\documentclass{intech}

\headertext{Extragalactic Compact Sources in the Planck Sky} 

\booktitle{Will-be-set-by-IN-TECH}%

\chaptertitle{Extragalactic Compact Sources\\in the Planck Sky \\ and their
Cosmological Implications}

\authors{Toffolatti Luigi*, Burigana Carlo
\\Arg\"{u}eso Francisco, and Diego Jos\'e M.}

\correspondingauthor{*Corresponding Author: ltoffolatti@uniovi.es\\
Full list of author information is available at the end of the chapter}

\dates{Received 25 August 2012; Accepted 29 August 2012\\
http://www.intechopen.com/10.5772/xxxxxxxx}

\begin{document}
\maketitle
\section*{Abstract}

The {\it Planck} satellite, launched into space on 14 May 2009, has proved to be a very successful space mission,
which has been operating flawlessly for more than 36 months. The main purpose of {\it Planck} is to map the
anisotropies of the cosmic microwave background (CMB) radiation on the whole sky at nine frequency channels,
between 30 and 857 GHz, with unprecedented resolution and sensitivity. After a brief description of the main
properties and cosmological aims of {\it Planck}, we review in this Chapter the most relevant results obtained by
{\it Planck} during the first 1.6 full-sky surveys relative to compact extragalactic sources and their
cosmological implications\footnote{ This paper is based largely on the {\it Planck} Early Release Compact Source
Catalogue, a product of ESA and the {\it Planck} Collaboration.  Any material presented in this review that is not
already described in {\it Planck} Collaboration papers represents the views of the authors and not necessarily
those of the {\it Planck} Collaboration.}. The most recent and efficient methods and filters for compact source
detection in the presence of Gaussian noise and of CMB anisotropies are discussed. Thanks to these methods, the
first full-sky surveys of {\it Planck} have enabled the detection of about two hundred of galaxy clusters, by the
Sunyaev--Zeldovich (SZ) effect, and from several hundreds to many thousands of extragalactic point sources (EPS),
all presented in the {\it Planck} Early Release Compact Source Catalogue. We review the latest results of {\it
Planck} on the SZ effect in galaxy clusters. We also discuss the possibilities of future {\it Planck} data to
study clusters through the SZ effect as well as the interesting possibilities of combining these data with X-ray
observations. The EPS observed by {\it Planck} can be classified into two main source classes: a) radio sources,
i.e., sources with emission dominated by non-thermal synchrotron radiation, at intermediate to high--redshift; b)
far--infrared (far-IR) sources, i.e. nearby dusty galaxies with a spectrum dominated by thermal dust emission,
typically at very low redshift. {\it Planck}'s number counts of EPS at LFI frequencies are in agreement with
Wilkinson Mircrowave Anisotropy Probe (WMAP) counts. However, a clear steepening of the mean spectral index of
bright radio sources is observed above 70 GHz, that can be interpreted in terms of a ``break'' frequency in the
spectra of sources. For nearby dusty galaxies, {\it Planck} observations find evidence of colder dust, with $T<
20$ K, than has previously been found. More recently, the number counts of bright local dusty galaxies and of
synchrotron sources have been measured by {\it Planck}, for the first time, at sub--millimetre wavelengths.
Finally, {\it Planck} has also provided interesting results on the angular distribution of Cosmic Infrared
Background (CIB) anisotropies, which allows us to put new constraints on clustering properties of dusty galaxies
at high redshift.

\section{Introduction}

\def\lsim{\,\lower2truept\hbox{${< \atop\hbox{\raise4truept\hbox{$\sim$}}}$}\,}
\def\gsim{\,\lower2truept\hbox{${> \atop\hbox{\raise4truept\hbox{$\sim$}}}$}\,}

As of mid August 2012, the {\it Planck} cosmic microwave background anisotropy probe \footnote{{\it Planck}
(http://www.esa.int/Planck) is a project of the European Space Agency - ESA - with instruments provided by two
scientific Consortia funded by ESA member states (in particular the lead countries: France and Italy) with
contributions from NASA (USA), and telescope reflectors provided in a collaboration between ESA and a scientific
Consortium led and funded by Denmark.} [1,2] -- launched into space on 14 May 2009 at 13:12:02 UTC, by an Ariane 5
ECA launcher, from the Guiana Space Centre, Kourou, French Guiana -- is still successfully operating. The
spacecraft accumulated data with its two instruments, the High Frequency Instrument (HFI) \cite{Planck_HFIa},
based on bolometers working between 100 and 857 GHz, and the Low Frequency Instrument (LFI) \cite{Planck_LFI},
based on radiometers working between 30 and 70 GHz, up to the consumption of the cryogenic liquids on January
2012, achieving $\simeq 29.5$ months of integration, corresponding to about five complete sky surveys. A further
12 months extension is on-going for observations with LFI only, cooled down with the cryogenic system provided by
HFI. Moreover, {\it Planck\/} is sensitive to linear polarization up to $353\,$GHz.


Thanks to its great sensitivity and resolution on the whole sky and to its wide frequency coverage that allows a
substantial progress in foreground modeling and removal, {\it Planck\/}  will open a new era in our understanding
of the Universe and of its astrophysical structures (see \cite{BlueBook2005} for a full description of the {\it
Planck\/} Scientific programme). {\it Planck\/} will improve the accuracy of current measures of a wide set of
cosmological parameters by a factor from $\sim 3$ to $\sim 10$ and will characterize the geometry of the Universe
with unprecedented accuracy.
{\it Planck\/} will shed light on many of the open issues in the connection
between the early stages of the Universe and the evolution of the cosmic structures, from the characterization of
primordial conditions and perturbations, to the late phases of cosmological reionization.

The {\it Planck\/} perspectives on some crucial selected topics linking cosmology to fundamental physics (the
neutrino masses and effective number of species, the primordial helium abundance, various physics fundamental
constants, the parity property of CMB maps and its connection with CPT symmetry with emphasis to the Cosmic
Birefringence, the detection of the stochastic field of gravitational waves) will also show how {\it Planck\/}
represents an extremely powerful {\it fundamental and particle physics laboratory}. Some of these analyses will be
carried out mainly through a precise measure of CMB anisotropy angular power spectrum (APS) in temperature,
polarization and in their correlations, whereas others, in particular those related to the geometry of the
Universe and to the research of non-Gaussianity signatures, are based on the exploitation of the anisotropy
pattern. The most ambitious goal is the possible detection of the so-called B-mode APS.

The first scientific results\footnote{
http://www.sciops.esa.int/index.php?project=PLANCK\&page=Planck\_Published\_Papers}, the so-called {\it Planck}
Early Papers \footnote{The {\it Planck} Early papers describe the instrument performance in flight including
thermal behaviour (papers I--IV), the LFI and HFI data analysis pipelines (papers V--VI), and the main
astrophysical results (papers VII-XXVI). These papers have complemented by a subsequent work, published in 2012,
based on a combination of high energy and {\it Planck} observations (see \cite{giommi12}).} have been released in
January 2011 and published by Astronomy and Astrophysics (EDP sciences), in the dedicated Volume 536 (December
2011). A further set of astrophysical results has been presented on the occasion of the Conference {\it
Astrophysics from radio to sub-millimeter wavelengths: the {\it Planck\/} view and other
experiments}\footnote{http://www.iasfbo.inaf.it/events/planck-2012/} held in Bologna on 13-17 February 2012.
Several articles have been already submitted in 2012 and others are in preparation, constituting the set of
so-called {\it Planck} Intermediate Papers.

The outline of this Chapter is as follows: in Section 2 we briefly sketch the main characteristics and the
capabilities of the ESA {\it Planck} mission; in Section 3 we discuss the most recent detection methods for
compact source detection; in Section 4 the SZ effect, detected by {\it Planck} in many cluster of galaxies and its
importance for cosmological studies are analyzed; Section 5 is dedicated to summarize current results obtained by
{\it Planck} data on the properties of EPS; finally, Section 6, discusses the very important results up to now
achieved by the analysis of CIB anisotropies detected by {\it Planck}.

\begin{table}[!ht]
  \caption{
{\it Planck\/} performance. The average sensitivity, $\delta$T/T, per (FWHM)$^2$ resolution element (FWHM: Full
Width at Half Maximum of the beam response function, is indicated in arcmin) is given in CMB temperature units
(i.e., equivalent thermodynamic temperature) for 29.5 (plus 12 for LFI) months of integration. The white noise
(per frequency channel for LFI and per detector for HFI) in 1~sec of integration (NET,  in  $\mu$K $\cdot
\sqrt{{\rm s}}$) is also given in CMB temperature units. The other acronyms here used are: 
N of R (or B) = number of radiometers (or bolometers), EB = effective bandwidth (in GHz). Adapted from
\cite{mandolesi10,lamarre10} and consistent with \cite{Planck_HFIa,Planck_LFI}. Note that at 100 GHz all
bolometers are polarized and the equivalent temperature value is obtained by combining polarization measurements.}

\begin{tabular}{l c c c}
& & & \\
\hline
LFI & & &\\
\hline
    Frequency (GHz) &   $30\,$  & $44\,$    & $70\,$ \\
\hline
    FWHM    & 33.34 &   26.81 & 13.03 \\
    N of R (or feeds)   & 4 (2) & 6 (3) & 12 (6) \\
    EB  & 6 & 8.8   & 14 \\
    NET & 159 & 197 & 158 \\
    $\delta$T/T [$\mu$K/K] (in $T$) & 2.04 & 3.14 & 5.17 \\
    $\delta$T/T [$\mu$K/K] (in $P$) & 2.88    & 4.44 & 7.32 \\
\hline
\end{tabular}
    \begin{tabular}{l c c}
& & \\

\hline
HFI & & \\
\hline
    Frequency (GHz)     & $100\,$   & $143\,$    \\
\hline
         FWHM  in $T$ ($P$)     & 9.6 (9.6) & 7.1 (6.9)  \\
            N of B in $T$ ($P$) & -- (8) & 4 (8)  \\
    EB  in $T$ ($P$) & 33 (33) & 43 (46)   \\
    NET in $T$ ($P$) & 100  (100) & 62 (82)  \\
    $\delta$T/T [$\mu$K/K] (in $T$) &  2.04 & 1.56 \\
    $\delta$T/T [$\mu$K/K] (in $P$) &  3.31 & 2.83 \\
\hline
\end{tabular}

\begin{tabular}{l c c c c}
\hline
HFI & & & & \\
\hline
    Frequency (GHz) &    $217\,$  & $ 353 \,$ & & \\
\hline
    FWHM     in $T$ ($P$) & 4.6 (4.6) &  4.7 (4.6) & &\\
    N of B in $T$ ($P$)   & 4 (8) & 4 (8) & &\\
    EB  in $T$ ($P$) &  72 (63)     & 99 (102) & &\\
    NET  in $T$ ($P$) & 91 (132) & 277 (404) & & \\
    $\delta$T/T [$\mu$K/K] in $T$ ($P$) & 3.31 (6.24)  & 13.7 (26.2) & & \\
\hline
\end{tabular}
    \begin{tabular}{l c c}
\hline
HFI & & \\
\hline 
    Frequency (GHz) &    $545\,$  & $ 857 \,$ \\
\hline
    FWHM     in $T$ & 4.7 &  4.3 \\
    N of B in $T$   & 4 & 4  \\
    EB  in $T$ & 169    & 257 \\
    NET in $T$ & 2000 & 91000 \\
    $\delta$T/T [$\mu$K/K] in $T$   & 103    & 4134 \\
\hline
\end{tabular}
\label{table:sens}
\end{table}


\section{The ESA {\it Planck} mission: overview}

CMB experimental data are affected by uncertainties due to instrumental noise (crucial at high multipoles, $\ell$,
i.e. small angular scales), cosmic and sampling variance (crucial at low $\ell$, i.e. large angular scales) and
from systematic effects. The uncertainty on the angular power spectrum is given by the combination of three
components, cosmic and sampling variance, and instrumental noise, and it is approximately given by \cite{knox95}:

\begin{equation}
\frac{\delta C_{\ell}}{C_{\ell}} = \sqrt{\frac{2}{f_{sky}(2 \ell +
1)}}  \left ( 1+ \frac{A \sigma^{2}}{N C_{\ell} W_{\ell}}  \right )
\, .
\end{equation}

\begin{figure}
\begin{minipage}[b]{0.3\linewidth}
\centering \vskip -0.2cm
\includegraphics[width=11cm]{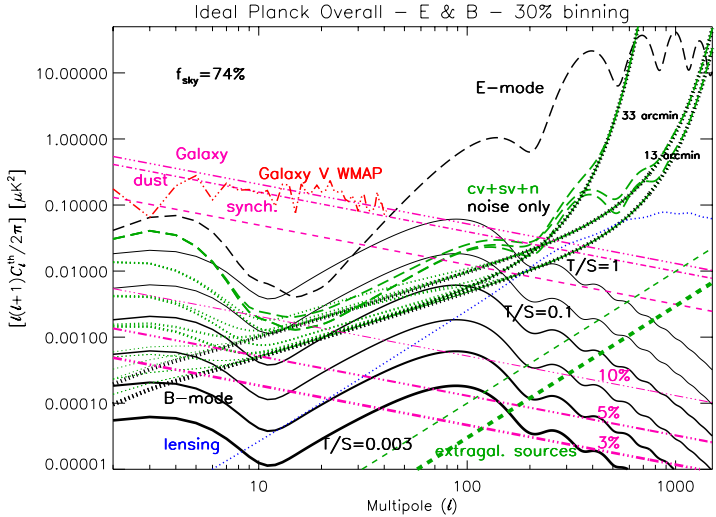}
\end{minipage}
\vskip -0.1cm

\caption{CMB E polarization modes (black long dashes) compatible with {\it WMAP\/} data and CMB B polarization
modes (black solid lines) for different tensor-to-scalar ratios $T/S=r$ of primordial perturbations are compared
to the {\it Planck\/} overall sensitivity to the APS assuming two different FWHM angular resolutions (33 and 13
arcmin) and the overall sensitivity corresponding to the whole mission duration (and also to two surveys only:
upper curve in black thick dots, labeled 13 arcmin). The expected noise is assumed to be properly subtracted. The
plots include cosmic and sampling variance plus instrumental noise (green dots for B modes, green long dashes for
E modes, labeled with cv+sv+n; black thick dots, noise only) assuming a multipole binning of 30\%.The B mode
induced by lensing (blue dots) is also shown. Galactic synchrotron (purple dashes) and dust (purple dot-dashes)
polarized emissions produce the overall Galactic foreground (purple three dots-dashes). {\it WMAP\/} 3-yr
power-law fits for uncorrelated dust and synchrotron have been used. For comparison, {\it WMAP\/} 3-yr results
(http://lambda.gsfc.nasa.gov/) derived from the foreground maps using HEALPix tools (http://healpix.jpl.nasa.gov/)
\cite{gorski05} are shown (red three dots-dashes broken line): power-law fits provide (generous) upper limits to
the power at low multipoles. Residual contamination levels by Galactic foregrounds (purple three dot-dashes) are
shown for 10\%, 5\%, and 3\% of the map level, at increasing thickness. We plot also as thick and thin green
dashes realistic estimates of the residual contribution of un-subtracted extragalactic sources, $C_\ell^{\rm
res,PS}$ and the corresponding uncertainty, $\delta C_\ell^{\rm res,PS}$.} \label{CMB_APS}

\end{figure}

\noindent Here $f_{sky}$ is the sky coverage, $A$ is the surveyed area, $\sigma$ is the instrumental rms noise per
pixel, $N$ is the total pixel number, $W_{\ell}$ is the beam window function that, in the case of a Gaussian
symmetric beam, is $W_{\ell} = {\rm{exp}} (-\ell (\ell + 1) \sigma_B^{2})$, with $\sigma_{B} = FWHM/\sqrt{8 \ln
2}$ the beamwidth which defines the angular resolution of the experiment. For $f_{sky} = 1$ the first term in
parenthesis defines the ``cosmic variance'', an intrinsic limit on the accuracy at which the APS of a certain
cosmological model defined by a suitable set of parameters can be derived with CMB anisotropy
measurements\footnote{Note that the cosmic and sampling  variance (74\% sky coverage excluding the sky regions
mostly affected by Galactic emission) implies a dependence of the overall sensitivity on $r$ at low multipoles,
relevant to the parameter estimation; instrumental noise only determines the capability of detecting the B mode.}.
It typically dominates the uncertainty on the APS at low $\ell$ because of the small, $2\ell + 1$, number of modes
$m$ for each $\ell$. The second term in parenthesis characterizes the instrumental noise, that never vanishes in
the case of real experiments. Note also the coupling between experiment sensitivity and resolution, the former
defining the low $\ell$ experimental uncertainty, namely for $W_{\ell}$ close to unit, the latter determining the
exponential loss in sensitivity at angular scales comparable with the beamwidth. We computed an overall
sensitivity value, weighted over the channels, defined by $1/\sigma_{j}^{2} = \sum_{i} 1/\sigma_{j,i}^{2}$, where
$j = T$ and $i$ indicates the sensitivity of each frequency channel, listed in Table \ref{table:sens}. FWHM values
of $13$ and $33$ arcmin are used in Fig.\ref{CMB_APS} to define the overall combination of {\it Planck}
sensitivity and resolution, i.e. the computation of the effective beam window function\footnote{In fact, it is
possible to smooth maps acquired at higher frequencies with smaller beamwidths to the lowest resolution
corresponding to a given experiment. We adopt here FWHM values of 33 and 13 arcmin, which correspond to the lowest
resolution of all the {\it Planck} instruments (i.e., 30 GHz channel) and to the lowest resolution of the so
called cosmological channels (i.e., 70 GHz channel), respectively (see Table 1).}, relevant for the sensitivity at
high $\ell$. Finally, to improve the signal to noise ratio in the APS sensitivity, especially at high multipoles,
a multipole binning is usually applied.
Of course, the real sensitivity of the whole mission will have to also include the potential residuals of
systematic effects. The {\it Planck} mission has been designed to suppress potential systematic effects down to
$\sim \mu K$ level or below. Fig.\ref{CMB_APS} compares CMB polarization modes with the ideal sensitivity of {\it
Planck\/} (including also a 15\% level of HFI data loss because of cosmic rays; see \cite{Planck_HFIb}) and the
signals coming from astrophysical foregrounds as discussed below.

CMB anisotropy maps are contaminated by a significant level of foreground emission of both Galactic and
extragalactic origin. For polarization, the most critical Galactic foregrounds are certainly synchrotron and
thermal dust emission, whereas free-free emission gives a negligible contribution. Other components, like spinning
dust and ``haze'', are still poorly known, particularly in polarization. Synchrotron emission is the dominant
Galactic foreground signal at low frequencies, up to $\sim$60 GHz, where dust emission starts to dominate.
External galaxies are critical only at high $\ell$, and extragalactic radio sources are likely the most crucial in
polarization up to frequencies $\sim$200 GHz, the most suitable for CMB anisotropy experiments. We parameterize a
potential residual from non perfect cleaning of CMB maps from Galactic foregrounds simply assuming that a certain
fraction of the foreground signal {\it at map level} (or, equivalently, its square at power spectrum level)
contaminates CMB maps. Of course, one can easily rescale the following results to any fraction of residual
foreground contamination. The frequency of 70 GHz, i.e. the {\it Planck} channel where Galactic foregrounds are
expected to be at their minimum level, at least at angular scales above $\sim $ one degree, is adopted as
reference.

\begin{figure}
\begin{minipage}[b]{0.3\linewidth}
\centering
\includegraphics[width=2.4cm,angle=90]{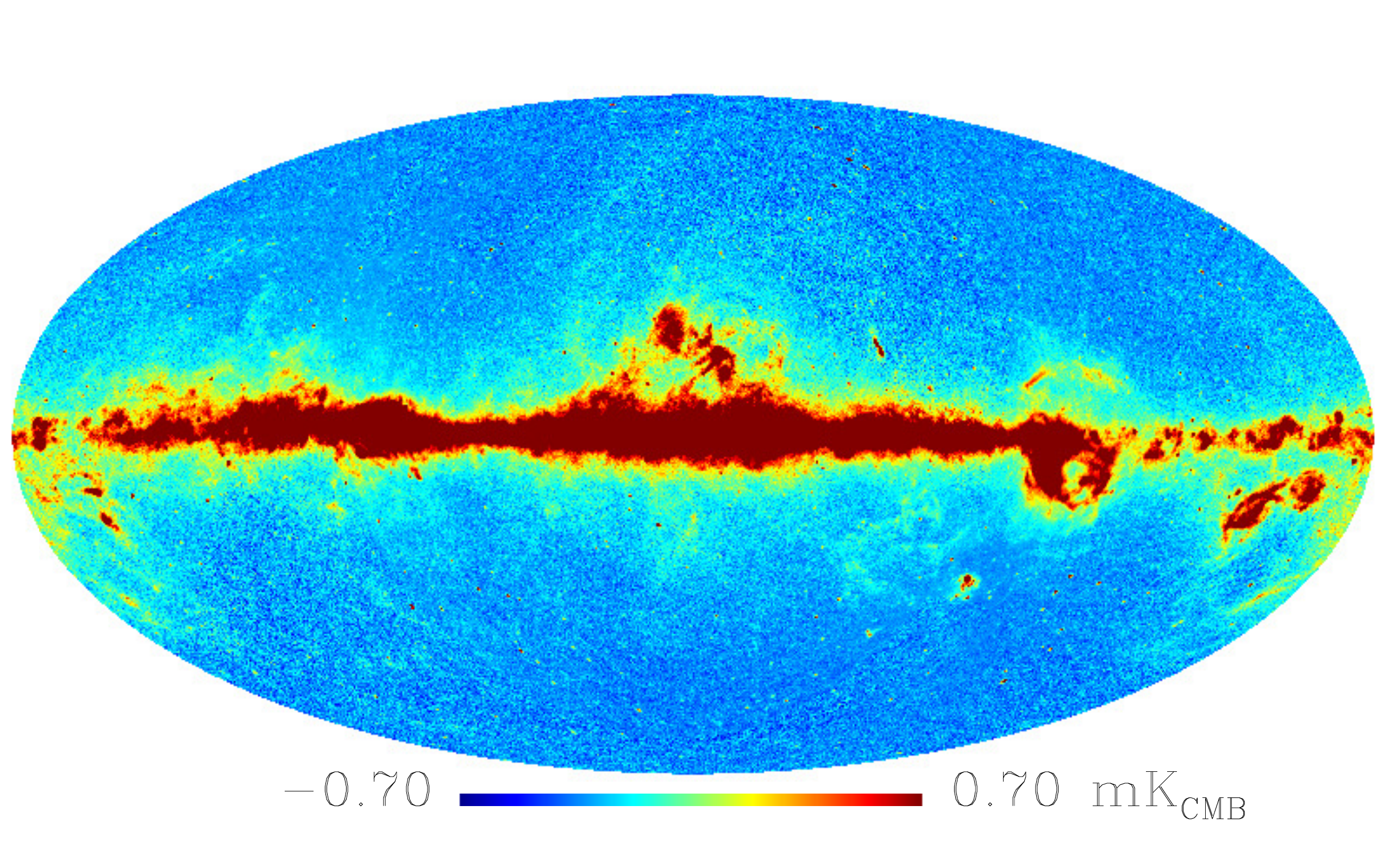}
\end{minipage}
\begin{minipage}[b]{0.3\linewidth}
\centering
\includegraphics[width=2.4cm,angle=90]{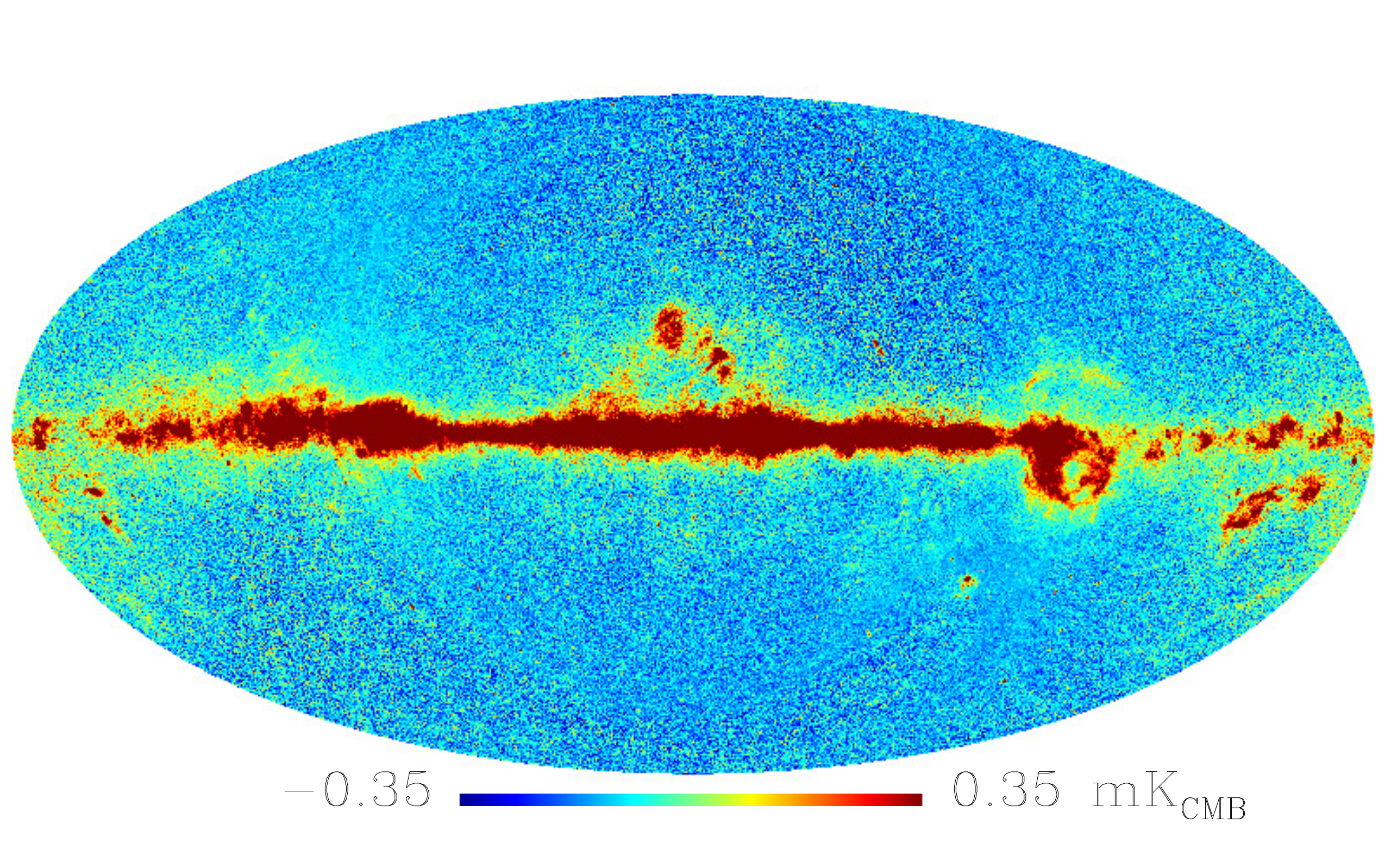}
\end{minipage}
\begin{minipage}[b]{0.3\linewidth}
\centering
\includegraphics[width=2.4cm,angle=90]{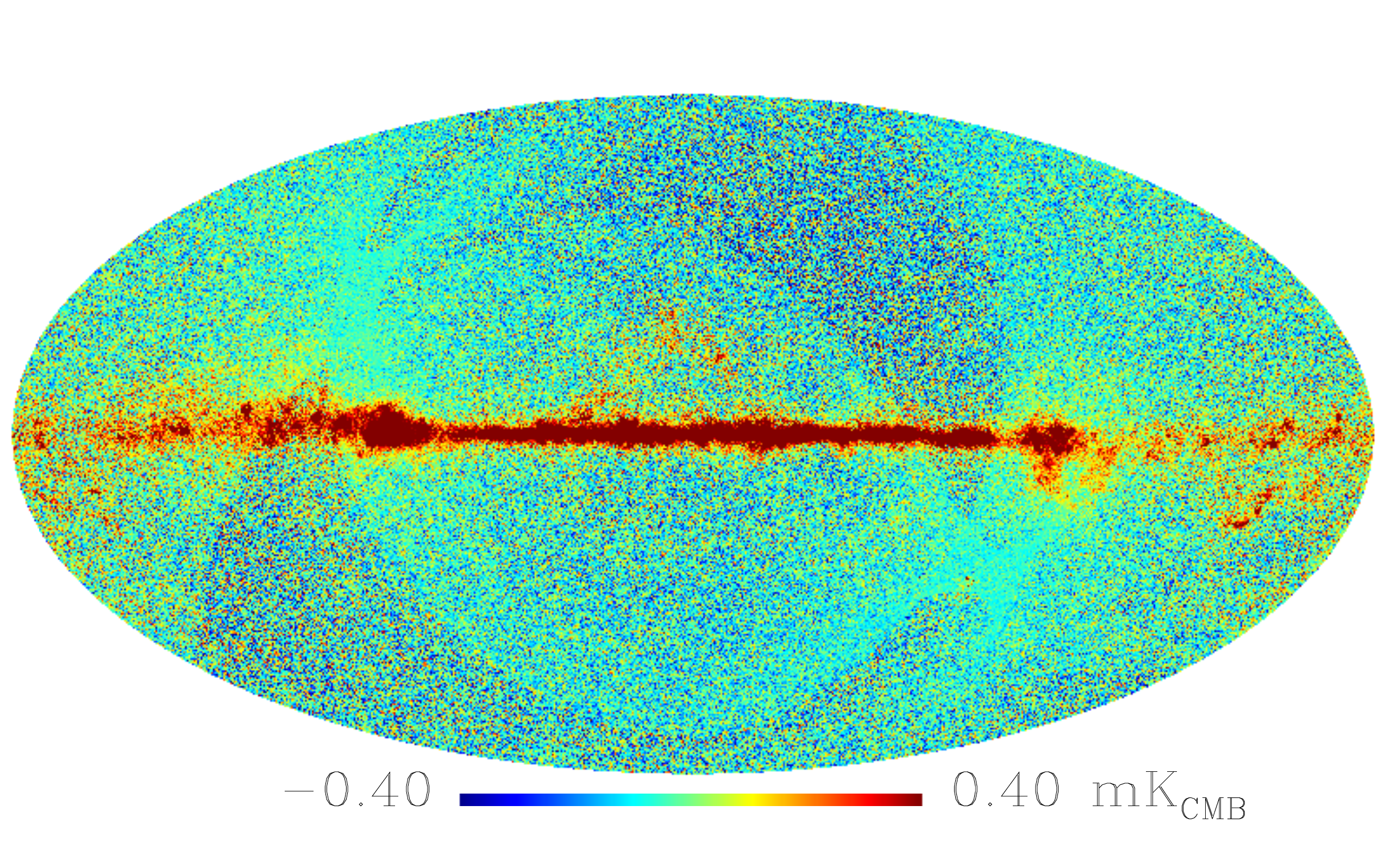}
\end{minipage}
%
\begin{minipage}[b]{0.3\linewidth}
\centering
\includegraphics[width=3.7cm]{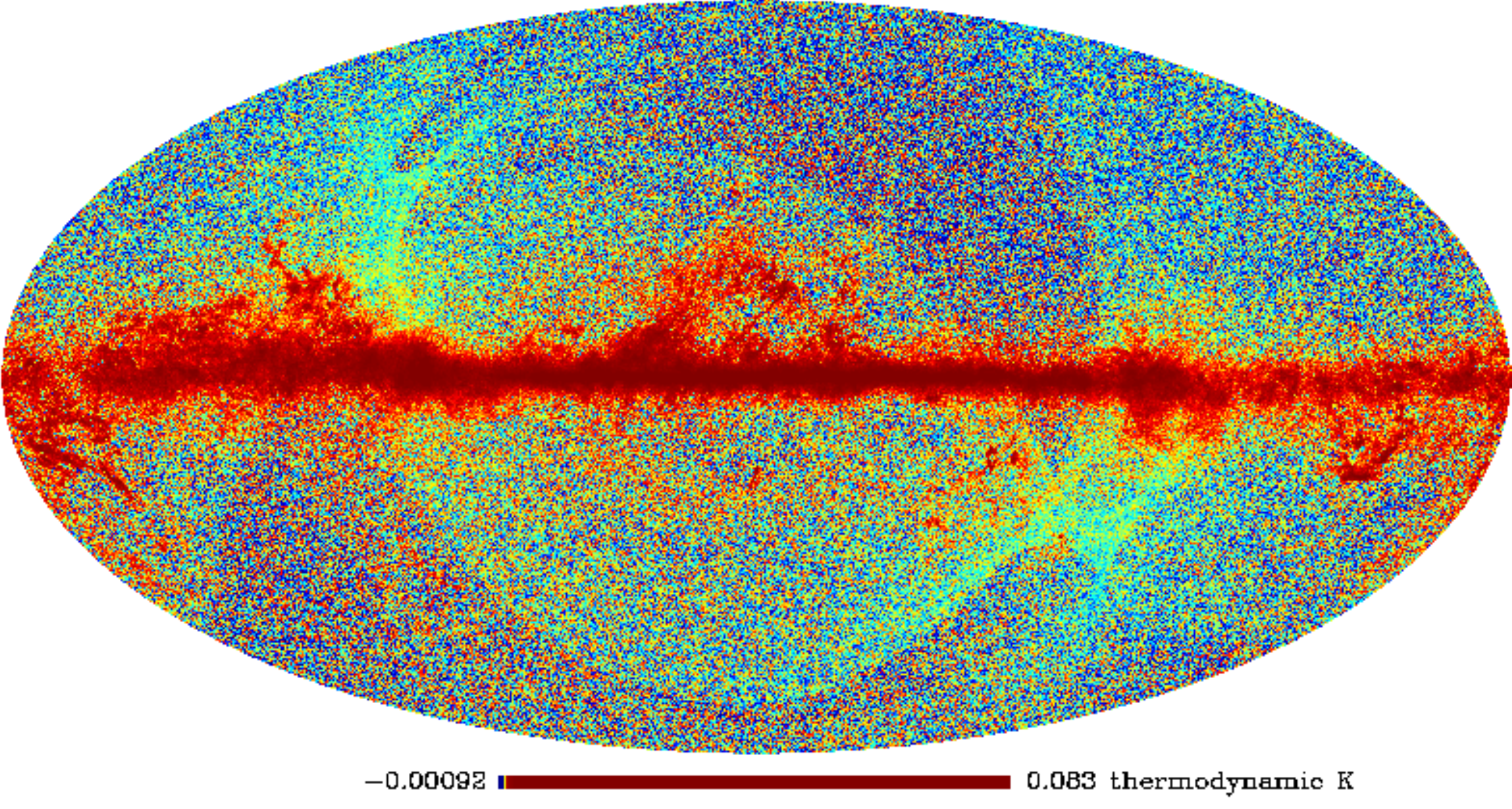}
\end{minipage}
\begin{minipage}[b]{0.3\linewidth}
\centering
\includegraphics[width=3.7cm]{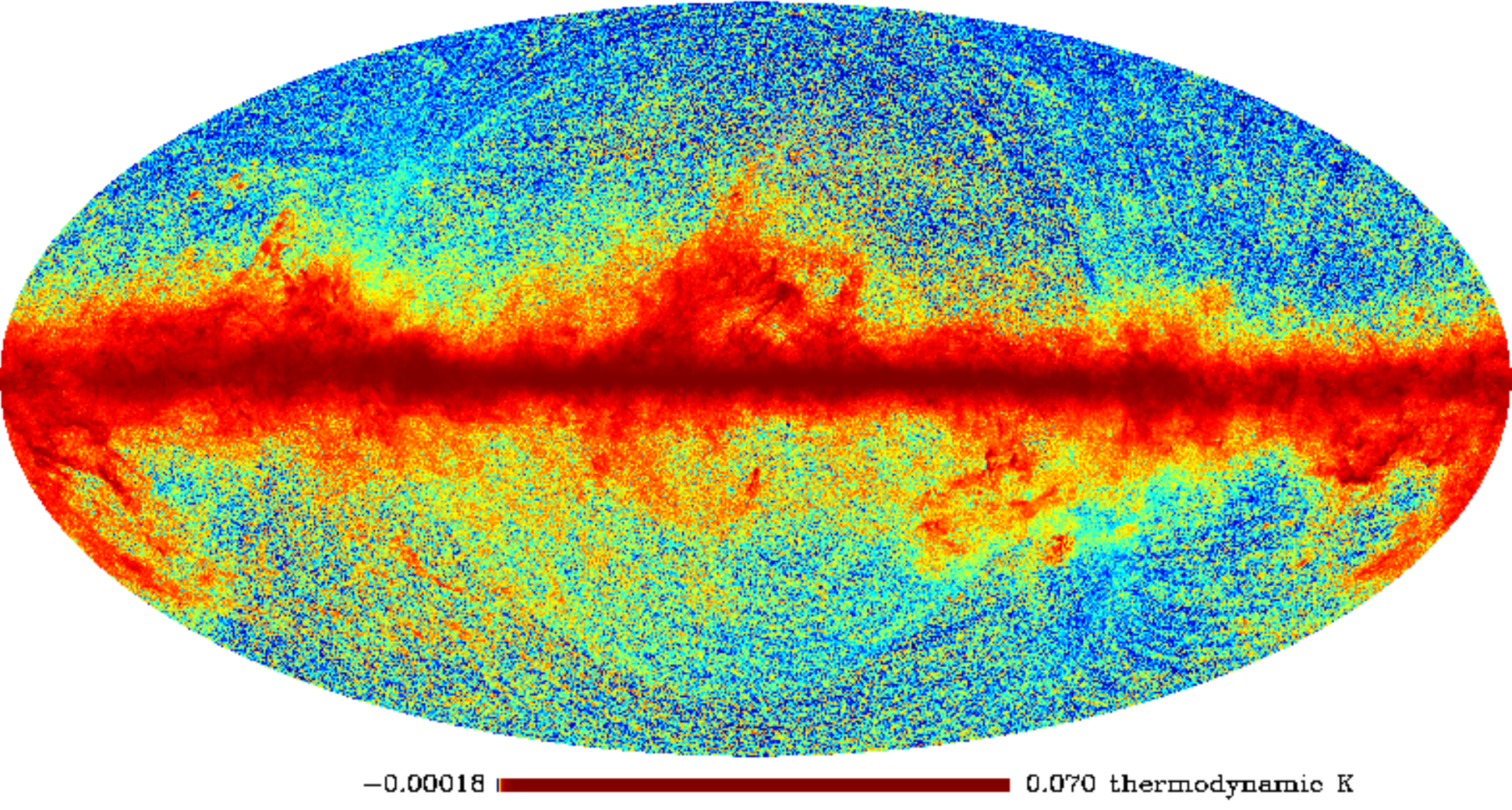}
\end{minipage}
\begin{minipage}[b]{0.3\linewidth}
\centering
\includegraphics[width=3.7cm]{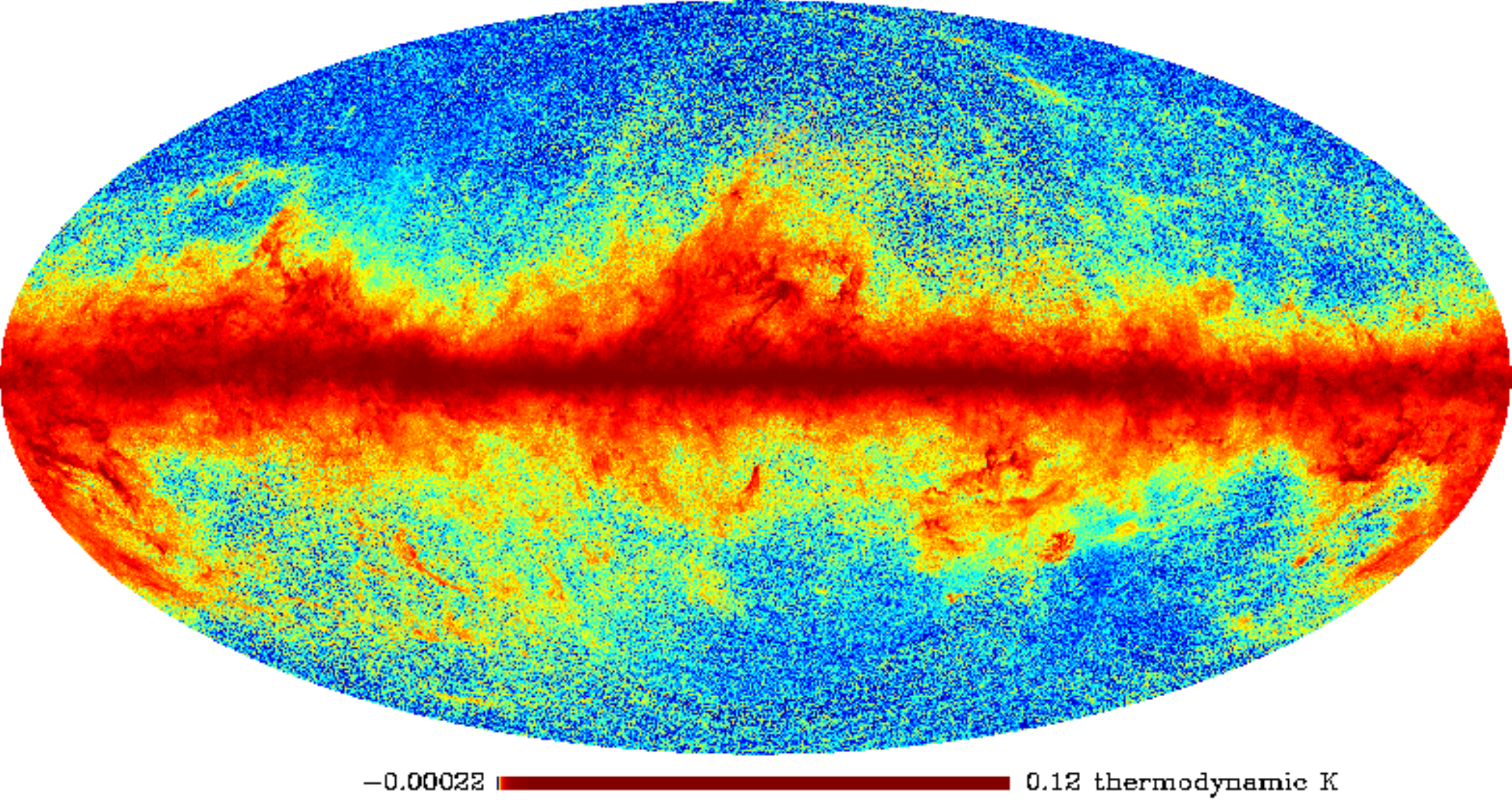}
\end{minipage}
%
\begin{minipage}[b]{0.3\linewidth}
\centering
\includegraphics[width=3.6cm]{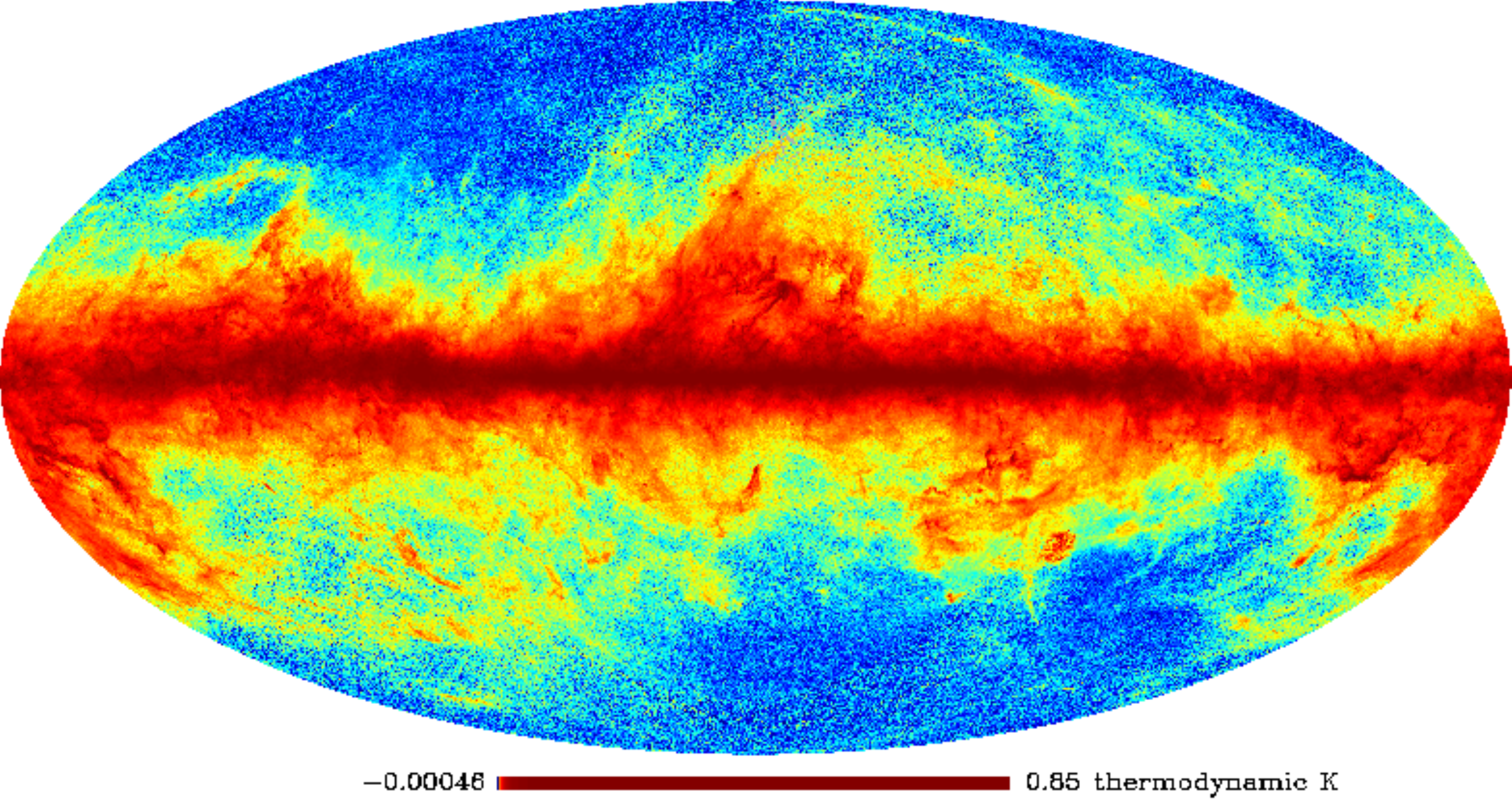}
\end{minipage}
\begin{minipage}[b]{0.36\linewidth}
\centering
\includegraphics[width=3.6cm]{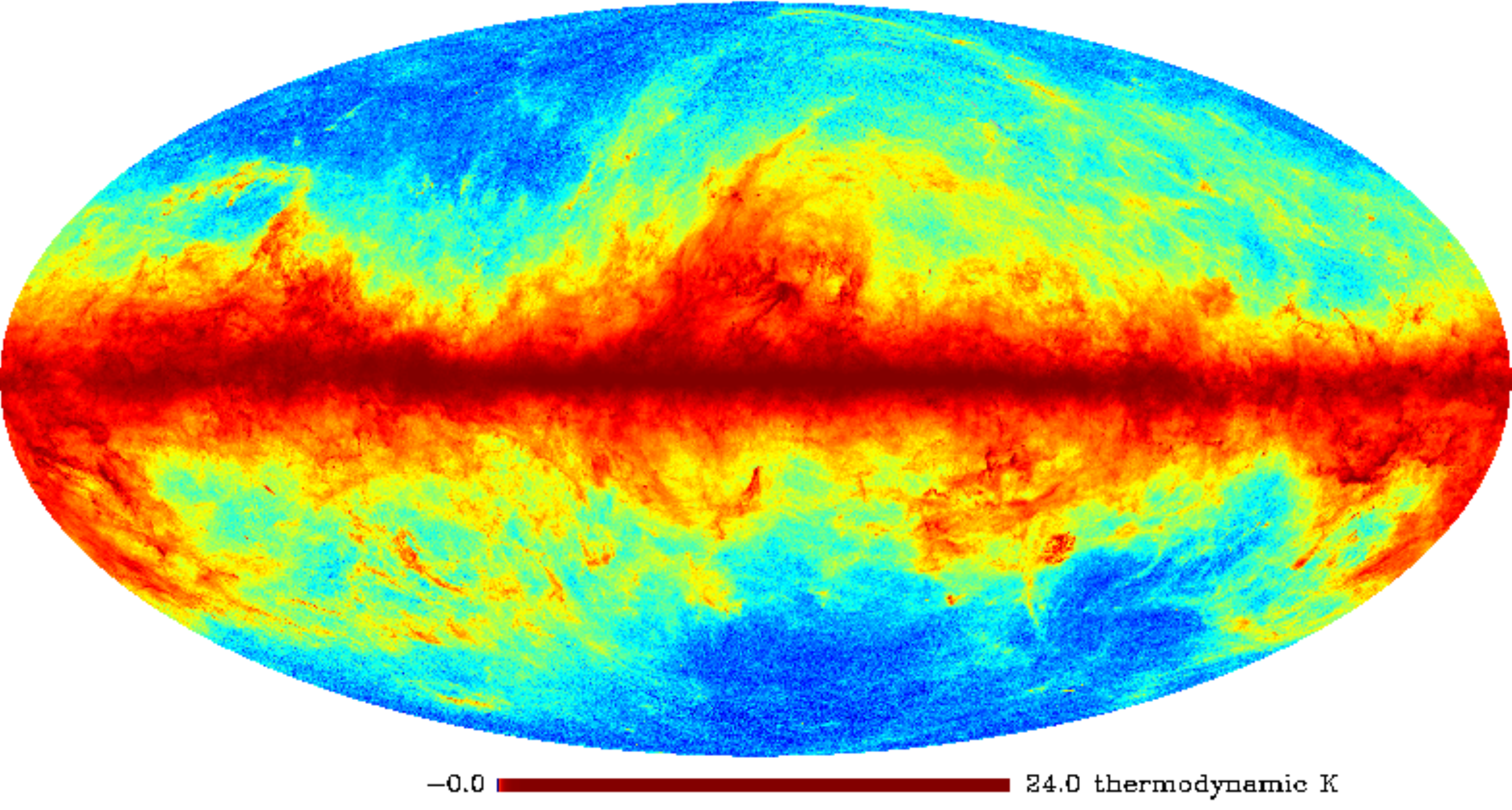}
\end{minipage}
\begin{minipage}[b]{0.36\linewidth}
\centering
\includegraphics[width=3.6cm]{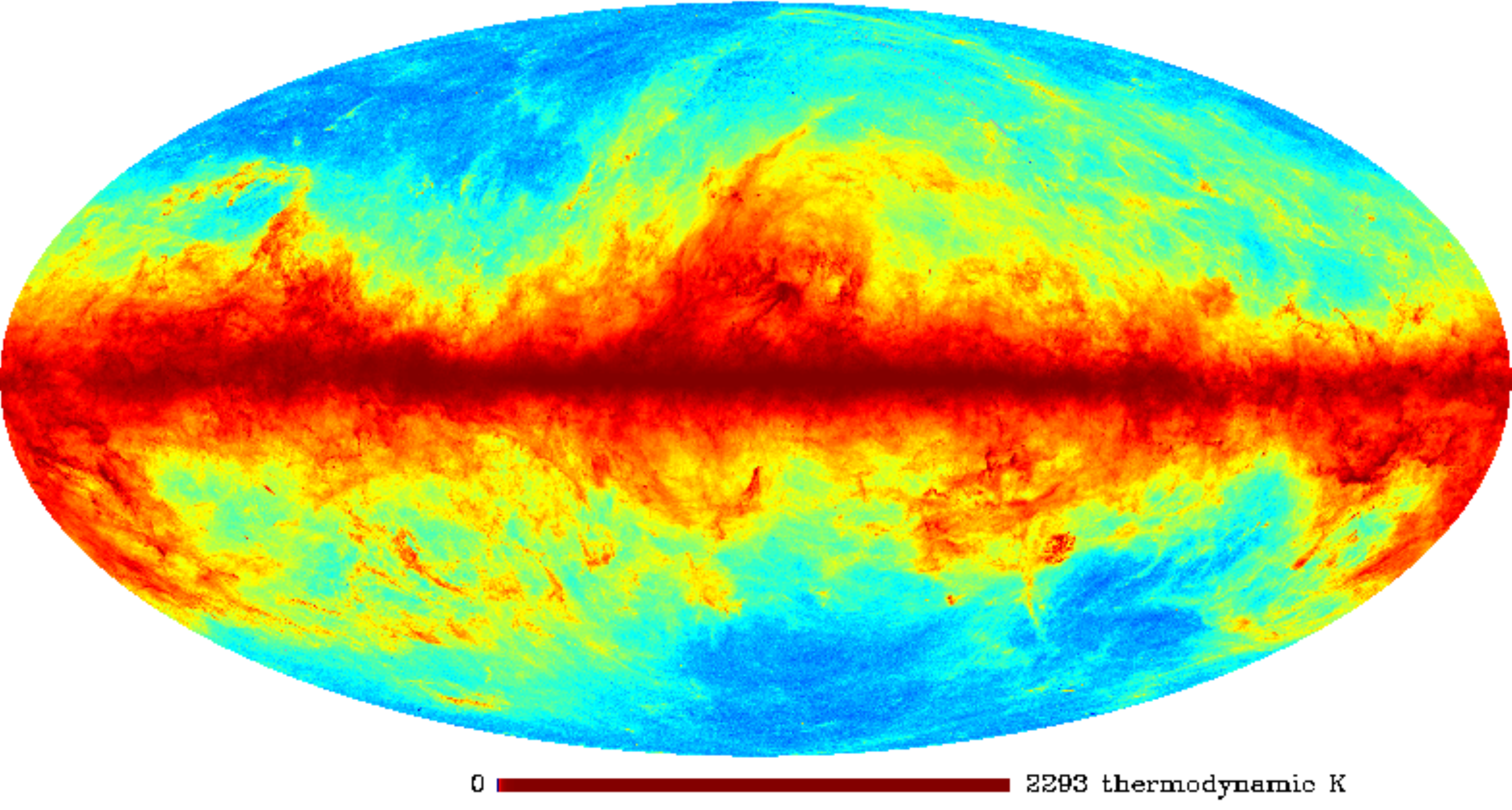}
\end{minipage}
\caption{CMB removed {\it Planck} full-sky maps. From left to right and from top to bottom: 30, 44, 70; 100, 143,
217; 353, 545, and 857 GHz, respectively. Credits: Zacchei, et al., A\&A, Vol. 536, A5, 2011; {\it Planck} HFI
Core Team, A\&A, Vol. 536, A6, 2011b, reproduced with permission by ESO.} \label{planck_freq_maps}
\end{figure}

For what concerns CMB temperature fluctuations produced by undetected EPS \cite{Toffolatti98}, we adopt the recent
(conservative) estimate of their Poisson contribution to the (polarized) APS \cite{Tucci-Toffolatti2012} at 100
GHz\footnote{We adopt here a frequency slightly larger than that considered for Galactic foregrounds (70 GHz)
because at small angular scales, where point sources are more critical, the minimum of foreground contamination is
likely shifted to higher frequencies.} by assuming a detection threshold of $\simeq 0.1$ Jy. We also assume a
potential residual coming from an uncertainty in the subtraction of this contribution computed by assuming a
relative uncertainty of $\simeq 10$\% in the knowledge of their degree of polarization and in the determination of
the source detection threshold, implying a reduction to $\simeq 30$\% of the original level. Except at very high
multipoles, their residual is likely significantly below that coming from Galactic foregrounds.

The first publications of the main cosmological (i.e. properly based on {\it Planck} CMB maps) implications are
expected in early 2013, together with the delivery of a first set of {\it Planck\/} maps and cosmological products
coming from the first 15 months of data. They will be mainly based on temperature data. Waiting for these
products, a first multifrequency view of the {\it Planck} astrophysical sky has been presented in the Early
Papers: Fig. \ref{planck_freq_maps} reports the first LFI and HFI frequency (CMB subtracted) maps. These maps are
the basis of the construction of the {\it Planck\/} Early Release Compact Source Catalog (ERCSC) (see
\cite{Planck_Paper7} and {\it The Explanatory Supplement to the Planck Early Release Compact Source Catalogue}),
the first {\it Planck} product delivered to the scientific community.

%

\begin{figure}
\begin{minipage}[b]{0.3\linewidth}
\centering \vskip -0.5cm
\includegraphics[width=11cm]{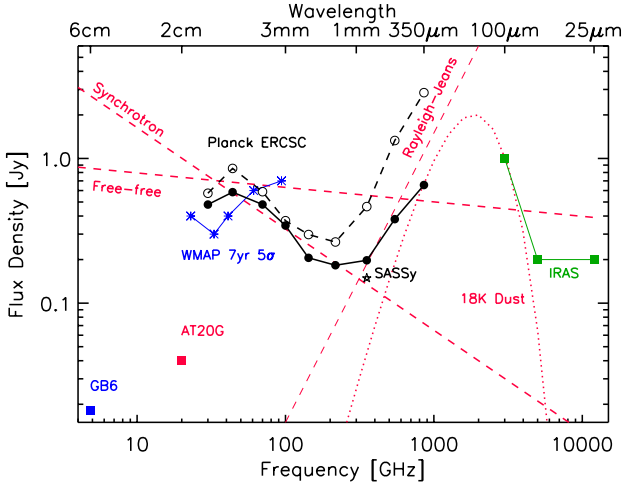}
\end{minipage}
\vskip -6.5cm \caption{The {\it Planck} ERCSC flux density limit quantified as the faintest ERCSC sources at
$\vert b\vert< 10^\circ$ (dashed black line) and at $\vert b\vert > 30^\circ$ (solid black line) is shown relative
to other wide area surveys. See Fig. 5 of \cite{Planck_Paper7} for more details. Credit: {\it Planck}
Collaboration, A\&A, vol. 536, A7, 2011, reproduced with permission by ESO.} \label{ercsc_sens}
\end{figure}

Fig. \ref{ercsc_sens} compares the sensitivity of {\it Planck} ERCSC with those of other surveys from radio to
far-infrared wavelengths. Of course, by accumulating sky surveys and refining data analysis, the {\it Planck\/}
sensitivity to point sources will significantly improve in the coming years. The forthcoming {\it Planck\/} Legacy
Catalog of Compact Sources (PCCS), to be released in early 2013 and to be updated in subsequent years, will
represent one of the major {\it Planck} products relevant for multi-frequency studies of compact or point--like
sources.

\subsection{Extragalactic point sources vs. non-Gaussianity}

The cosmological evolution of extragalactic sources and its implications for the CMB and the CIB will be discussed
in the following Sections, 5 and 6. On the other hand, statistical analyses of the extragalactic source
distribution in the sky can be applied to test cosmological models. In this context, the possibility of probing
the Gaussianity of primordial perturbations appears particularly promising. Primordial perturbations at the origin
of the large scale structure (LSS) may leave their imprint in the form of small deviations from a Gaussian
distribution, \cite{Komatsu10,Bartolo04} that can appear in different kinds of configurations, such as the
so-called local type, equilateral, enfolded, orthogonal. For example, the local type of deviation from Gaussianity
is parameterized by a constant dimensionless parameter $f_{NL}$ (see, e.g., \cite{Verde00,Komatsu01,Babich04})
 $\Phi = \phi + f_{NL} ({\phi}^2-\left\langle{\phi}^2\right\rangle)$, where $\Phi$ denotes
Bardeen's gauge-invariant potential (evaluated deep in the matter era in the CMB convention) and $\phi$ is a
Gaussian random field. Extragalactic radio sources are particularly interesting as tracers of the LSS since they
span large volumes extending out to substantial redshifts. The radio sources from the NRAO VLA Sky Survey (NVSS),
the quasar catalogue of Sloan Digital Sky Survey Release Six (SDSS DR6 QSOs) and the MegaZ-LRG (DR7), the final
SDSS II Luminous Red Galaxy (LRG) photometric redshift survey, have been recently analysed by \cite{Xia11} (see
this work and references therein for a thorough analysis on the subject). Through a global analysis of the
constraints on the amplitude of primordial non-Gaussianity by the angular power spectra obtained from
extragalactic radio sources (mapped by these surveys) and, moreover, from their cross-correlation power spectra
with the WMAP CMB temperature map, \cite{Xia11} set limits on $f_{NL}=48\pm20$, $f_{NL} =50\pm265$ and
$f_{NL}=183\pm 95$ at 68\% confidence level for local, equilateral and enfolded templates, respectively. These
results have been found to be stable with respect to potential systematic errors: the source number density and
the contamination by Galactic emissions, for NVSS sources; the use of different CMB temperature fluctuation
templates and the contamination of stars in the SDSS and LRG samples. Such tests of non--Gaussianity would have
profound implications for inflationary mechanisms -- such as single-field slow roll, multifields, curvaton (local
type) -- and for models which have effects on the halo clustering can be described by the equilateral template
(related to higher-order derivative type non-Gaussianity) and by the enfolded template (related to modified
initial state or higher-derivative interactions). Fundamental progress on this topic will be achieved by combining
forthcoming LSS surveys with the CMB maps provided by {\it Planck}.

\section{Methods for compact source detection in CMB maps}

Compact sources, in CMB literature, are defined as spatially bounded sources which subtend very small angular
scales in the images, such as galaxies and galaxy clusters. On the other hand, diffuse components, such as the CMB
itself and Galactic foregrounds, do not show clear spatial boundaries and extend over large areas of the sky. Due
to the fact that compact sources are spatially localized, the techniques for detecting them differ from those
applied for the separation of the diffuse components. Most of the detection methods use scale diversity, i.e.
different power at different angular scales, to enhance compact sources against diffuse components. Sources must
be detected against a combination of CMB, instrumental noise and Galactic foregrounds. From the point of view of
signal processing, the source is the signal and the other components are the noise.

Point sources are ``compact'' sources in the sense that their typical observed angular size is much smaller than
the beam resolution of the experiment. Therefore, they appear as point-like objects convolved with the
instrumental beam. Radio sources and far--IR sources are usually seen as point-like sources. Galaxy clusters,
which are detected through the thermal SZ effect \cite{SZE}, have a shape that is obtained as the convolution of
the instrumental beam with the cluster profile. In contrast to point sources, the cluster profile has to be taken
into account for cluster detection. However, since the projected angular scale of clusters is generally small,
techniques that are useful for point sources can be adapted for clusters, too.

The thermal SZ effect has a general dependence with frequency, that makes the use of multichannel images very
convenient for cluster detection. On the contrary, the flux of each individual point source has its own frequency
dependence. Despite this fact, the combination of several channels can also improve point source detection. We
will review techniques applied to single-frequency channels in a first subsection and then we will discuss more
recent methods, that use multichannel information.

\subsection{Single channel detection}

We now focus on techniques for detecting point--like sources. Galaxy clusters can be detected by similar methods,
but taking into account the cluster profile (see, e.g., [23-25]). Since multichannel methods for cluster detection
improve significantly the performance of single channel techniques, we leave a more detailed study of clusters for
the next subsection.

As mentioned before, compact source detection techniques make use of scale diversity. For example, SEXTRACTOR
\cite{Bertin96} -- where maps are pre--filtered by a Gaussian kernel the same size as the beam -- approximates the
image background by a low-order polynomial and then subtracts the background from the image. The object is
detected after connecting the pixels above a given flux density threshold. SEXTRACTOR has been used for
elaborating the {\it Planck} ERCSC \cite{Planck_Paper7} in the highest frequency channels, from 217 to 857 GHz.
However, CMB emission and diffuse foregrounds are complex and cannot be modeled in a straightforward way. Thus,
apart from this important application, SEXTRACTOR has had a limited use in CMB astronomy.

A standard method which has been used often for compact source detection is the common matched filter (MF) [27].
The MF is just a linear filter with suitable characteristics for amplifying the source against the background. The
image $y(\vec{x})$ is convolved with a filter $\psi(\vec{x})$:
\begin{equation}
 \omega(\vec{x})=\int  y(\vec{u})\psi(\vec{x}-\vec{u})\, d\vec{u}
\end{equation}

The MF is defined as the linear filter that is an unbiased estimator of the source flux and minimizes the variance
of the filtered map. In order to satisfy these mathematical constraints, if we assume that the beam is circularly
symmetric and the background a homogeneous and isotropic random field, the MF must be defined in Fourier space as
\begin{equation}
\psi(q)=k \displaystyle\frac{\tau(q)}{P(q)}
\end{equation}

\noindent where $\tau(q)$ is the Fourier transform of the beam, $P(q)$ the background power spectrum and k the
normalization constant. With this definition, the MF gives the maximum amplification of the compact source with
respect to the background. Once the source has been amplified, point sources are detected in the filtered map as
peaks above a given threshold, typically$5\sigma$, with $\sigma$ the r.m.s deviation of the
filtered map. The MF has been used both with simulations [28] 
and real data [29]. In this last paper, the WMAP team estimated the fluxes by fitting to a Gaussian profile plus a
planar baseline. In [24] the MF was also applied to the detection of clusters.
The use of wavelets for source detection is an interesting alternative to the MF. Wavelets are compensated linear
filters, i.e their integral is zero, that help to remove the background contribution and yield a high source
amplification. Since the beam is approximately Gaussian, the Mexican Hat Wavelet (MHW), constructed as the
Laplacian of a Gaussian function, adapts itself very well to the detection problem. The MHW depends on the scale
$R$, a parameter that determines the width of the wavelet:
\begin{equation}
\psi(q)\propto (qR)^2 \exp \big(-\frac{(qR)^2}{2}\big)
\end{equation}

The MHW has been succesfully applied to simulated CMB data [30]. The scale $R$ is fixed in order to obtain the
maximum amplification and the determination of the power spectrum is not necessary. A family of wavelets that
generalize the MHW, the Mexican Hat Wavelet Family (MHWF) was presented in [31]. The performance of this family
was compared with the MF in [28] and it produced similar results when implemented on {\it Planck} simulations. The MHWF
was also applied to point source detection in WMAP images [32] by using a non-blind method. This method yielded a
larger number of detections than the MF used by the WMAP group. The general expression of the MHWF is:

\begin{equation}
\psi(q)\propto (qR)^{2n} \exp \big(-\frac{(qR)^2}{2}\big)
\end{equation}

\noindent $n$ being a natural number. Further improvements can be obtained if we admit
any real exponent such as in the Bi-parametric Adaptive Filter (BAF) [33]. 

The MF and the diverse types of wavelets do not use any prior information about the average number of sources in
the surveyed patch, the flux distribution of the sources or other properties. Therefore, useful information is
being neglected by these methods. In contrast, Bayesian methods provide a natural way to incorporate information
about the statistical properties of both the source and the noise. Several Bayesian methods have been proposed in
the literature for the detection problem [24, 34, 35]. These methods construct a posterior probability $
Pr({\theta}|D,H) $ by using Bayes' theorem

\begin{equation}
   Pr(\theta|D,H)=\displaystyle\frac{Pr(D|\theta,H)Pr(\theta|H)}{Pr(D|H)}
\end{equation}

\noindent where $\theta$ are the relevant parameters (positions, fluxes, sizes, etc.), $D$ the data, and $H$ the
underlying hypothesis. In Bayesian terminology  $Pr(D|\theta,H)$ is the likelihood, $Pr(\theta|H)$ is the prior
and $Pr(D|H)$ is the Bayesian evidence. Different Bayesian techniques can differ in the priors or in the way of
exploring the complicated posterior probability. PowellSnakes [34] is an interesting method, which has been
applied with success to the compilation of the ERCSC for {\it Planck} frequencies between 30 and 143 GHz
\cite{Planck_Paper7}. It uses Powell's minimization and physically motivated priors. This method can be also
applied to cluster detection just by introducing a suitable prior on the cluster size.

A simple Bayesian way of determining the position of the sources and estimating their number and flux densities
has been presented in \cite{arg11}. Whereas by the MF, or by wavelets, sources are detected above an arbitrary
threshold, Bayesian methods select them in a more natural way, for instance by comparing the posterior probability
of two hypothesis: presence or absence of the source. In the next subsection we will explore multichannel methods
that help improve the detection performance

\subsection{Multi-channel detection}


The flux density distribution, $f_{\nu}$, of extragalactic radio sources as a function of frequency, $\nu$, is
usually approximated by a power law, although this approximation is only valid in limited frequency intervals,
i.e. $f_{\nu}\propto({\nu}/{\nu_0})^{\alpha}$, with $\nu_0$ being some frequency of reference. Nevertheless, the
so called ``spectral index'', $\alpha$, changes from source to source and this formula is not reliable when the
range of frequencies is wide enough. In \cite{herr08} a scheme for channel combination was proposed that makes the
spectral behavior irrelevant. This method is called matrix multifilters (MTXFs) and relies on the application of a
set ($N\times N $ matrix) of linear filters which combine the information of the $N$ channels in such a way that:
1) an unbiased estimator of the source flux at each channel is obtained; and 2) the variance of the estimator is
minimum. Note that the method does not mix the images in a single map, but it produces $N$ maps in which the
sources are conveniently amplified. The method defaults to the MF when there is no cross-correlation among the
channels. When there is a non negligible correlation among the channels, as is the case for microwave images taken
at different frequencies where CMB and Galactic foregrounds are present in all the images, this method gives a
clear increase of the amplification when compared with the MF.

Now, we discuss a method tailored for cluster detection through the thermal SZ effect. Matched Multifilters (MMF)
\cite{herr02b} combine $N$ channels in a single image, incorporating the information about the spectral behavior
(thermal SZ effect) and with the $N$ filters depending on a scale parameter $S$. The filters are constructed in
the usual way, by imposing unbiasedness and minimum variance. The MMF is given (in matrix notation) by

\begin{equation}
\textbf{$\Upsilon$}(q)=\alpha \textbf{P}^{-1}F, \,\,\, \alpha^{-1}=\int d\textbf{q}\, F^t \textbf{P}^{-1}F,
\end{equation}

\noindent where $F$ is the column vector $[f_{\nu}\tau_{\nu}] $, which incorporates the spectral behavior
$f_{\nu}$ and the shape of the cluster at each frequency $\tau_{\nu}$ and $\bf{P}$ is the cross-power spectrum.
Since the cluster size is not known a priori, the images are convolved with a set of filters with different scales
$S_i$, and it has been proven that the amplification is maximum when the chosen scale coincides with the cluster
size. A common pressure profile is assumed for the clusters. The detection is performed by searching for the
maxima of the filtered map above a given threshold. The estimated amplitude of the thermal SZ effect is given by
the amplitude at the maxima.

MMF can also be adapted to detect the fainter kinematic SZ effect. In this case an orthogonality condition with
respect to the spectral behavior of the thermal SZ effect is imposed. Together with the usual unbiasedness and
minimum variance conditions, this last constraint helps cancel out the thermal SZ effect contamination
\cite{herr05}. A MMF can also be designed for point source detection, by incorporating the (unknown) spectral
behavior of the sources as a set of parameters in the filter, it has also been proven that the amplification is
maximum when these parameters coincide with the real source spectrum. By changing the parameters and selecting
those which give the maximum amplification, in \cite{lanz12} the authors were able to increase the number of point
source detections in the WMAP 7-year maps.

Finally, a multi-channel Bayesian method has been developed recently, Powell-Snakes II \cite{carva12}. This method
constructs a posterior distribution by combining the likelihood and the prior information of the different
channels. It is an extension of Powell-Snakes I and uses prior information on the positions, number of sources,
intensities, sizes and spectral parameters. The method is suitable both for point sources and for clusters. It is
worth noting that maximizing the likelihood when the sources are well separated, i.e. in the absence of  source
blending, amounts to using the MMF presented above. Here we have briefly summarized the most important topics on
the subject: a more detailed review can be found in \cite{herr12}.

\section{Sunyaev-Zeldovich effect in clusters of galaxies}
\label{SZE} The Sunyaev-Zeldovich effect (SZ, \cite{SZE}) accounts for the interaction between a hot plasma (in a
cluster environment) and the photons of the CMB. When CMB photons cross a galaxy cluster, some of these photons
interact with the free electrons in the hot plasma through inverse Compton scattering. The temperature change
observed in a given direction, $\theta$, and at the frequency $\nu$, can be described as
\begin{equation}
\Delta T(\theta,\nu) = C_0 \int n_e(l) T(l) dl \label{eq_DeltaT}
\end{equation}
where $C_0$ contains all the relevant constants including the frequency dependence ($g_x = x(e^x + 1)/(e^x - 1) -
4$, with $x=h\nu/kT$), $n_e$ is the electron density and $T$ is the electron temperature. The integral is
performed along the line of sight.

The same electrons that interact with the CMB photons emit X-rays through the bremstrahlung process:
\begin{equation}
S_x(\theta,\nu) = S_0 \frac{\int n_e^2 T^{1/2} dl}{D_\ell(z)^2} \label{eq_Sx}
\end{equation}
where $D_\ell(z)$ is the luminosity distance. The quantity $S_0$ contains all the relevant constants and frequency
dependence. Combining X--ray and SZ observations it is thus possible to reduce the degeneracy between different
models due to their different dependency on $T$ and especially with $n_e$.

Due to the nature of the microwave radiation, water vapour (and hence our atmosphere) presents a challenge for
studying this radiation from the ground. Observations have to be carried out through several windows where the
transmission of the microwave light is maximized. In recent years, ground--based experiments have benefited from
important progress in the development of very sensitive bolometers. These bolometer arrays when combined with
superb atmospheric conditions -- found in places like the South Pole and the Atacama desert (with extremely low
levels of water vapour) -- have allowed, for the first time at all, galaxy clusters to be mapped in great detail
through the SZ effect. The South Pole Telescope (or SPT; see, e.g., \cite{hall10}) and the Atacama Cosmology
Telescope (or ACT; see, e.g., \cite{fowler10}) are today the most important ground--based experiments carrying out
these observations.

From space, the {\it Planck} satellite -- even though it lacks the spatial resolution of ground--based experiments
-- complements them by applying a full--sky coverage, a wider frequency range and a better understanding of
Galactic and extragalactic foregrounds. In particular, {\it Planck} is better suited than ground--based
experiments to detect large angular scale SZ signals like nearby galaxy clusters or the diffuse SZ effect. In
fact, ground--based experiments can have their large angular scales affected by atmospheric fluctuations that need
to be removed, carefully. This removal process can distort the modes that include the large angular scales signal.
On the contrary, {\it Planck} data does not suffer from these limitations and its relatively poor angular
resolution (if compared to some ground experiments) can be used to its advantage. The wide frequency coverage and
extremely high sensitivity of {\it Planck} allows for detailed foreground (and CMB) removal that could overwhelm
the weak signal of the SZ effect. {\it Planck} data will help improve the understanding of the distribution and
the characteristics of the plasma in clusters. The conclusions derived on the internal structure of clusters will
ultimately have an impact on other works that focus on deriving cosmological parameters. In fact, cosmological
studies cannot by themselves disentangle among the uncertainties in the physics inside galaxy clusters.

\begin{figure}[htb]
\begin{center}
\includegraphics[width=.5\textwidth]{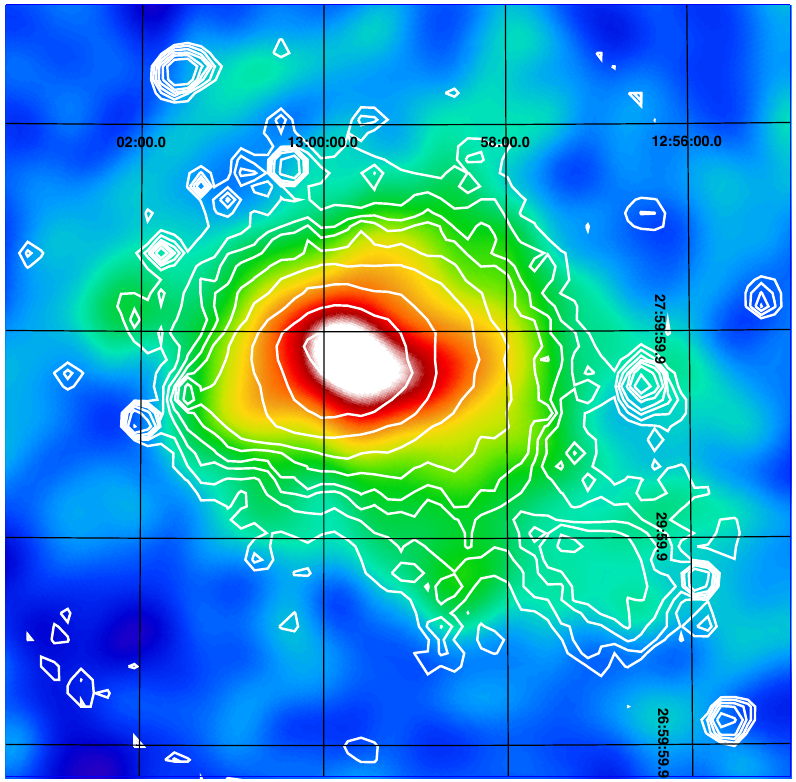}
\end{center}
\caption{Fig. 1 from \cite{PlanckClusterCatalog_Paper8}. Coma cluster as seen by {\it Planck}. Contours show the
the X-ray emission from Coma. Credit: {\it Planck} Collaboration, A\&A, Vol. 536, A8, 2011, reproduced with
permission by ESO.} \label{F1}
\end{figure}

The {\it Planck} satellite is currently detecting hundreds of clusters of galaxies through the thermal SZ effect.
One of the peculiarities of the SZ effect is that the change in the CMB temperature in the direction of a cluster
is independent of the distance to that cluster. This makes the SZ an ideal tool to explore the high redshift
Universe. {\it Planck} is perfect for studying the most massive clusters in the Universe and is expected to see
clusters beyond $z=1$. Earlier results on galaxy clusters obtained by {\it Planck} have been presented in a subset
of the {\it Planck} Early results papers and, more specifically, can be found in
\cite{PlanckClusterCatalog_Paper8,PlanckXMM_Paper9, Planck_Paper10,Planck_Paper11,Planck_Paper12,Planck_Paper26}
and also in \cite{Planck_paperWL}, where new results based on additional data are starting to be presented.

\begin{figure}[htb]
\begin{center}
\includegraphics[width=.7\textwidth]{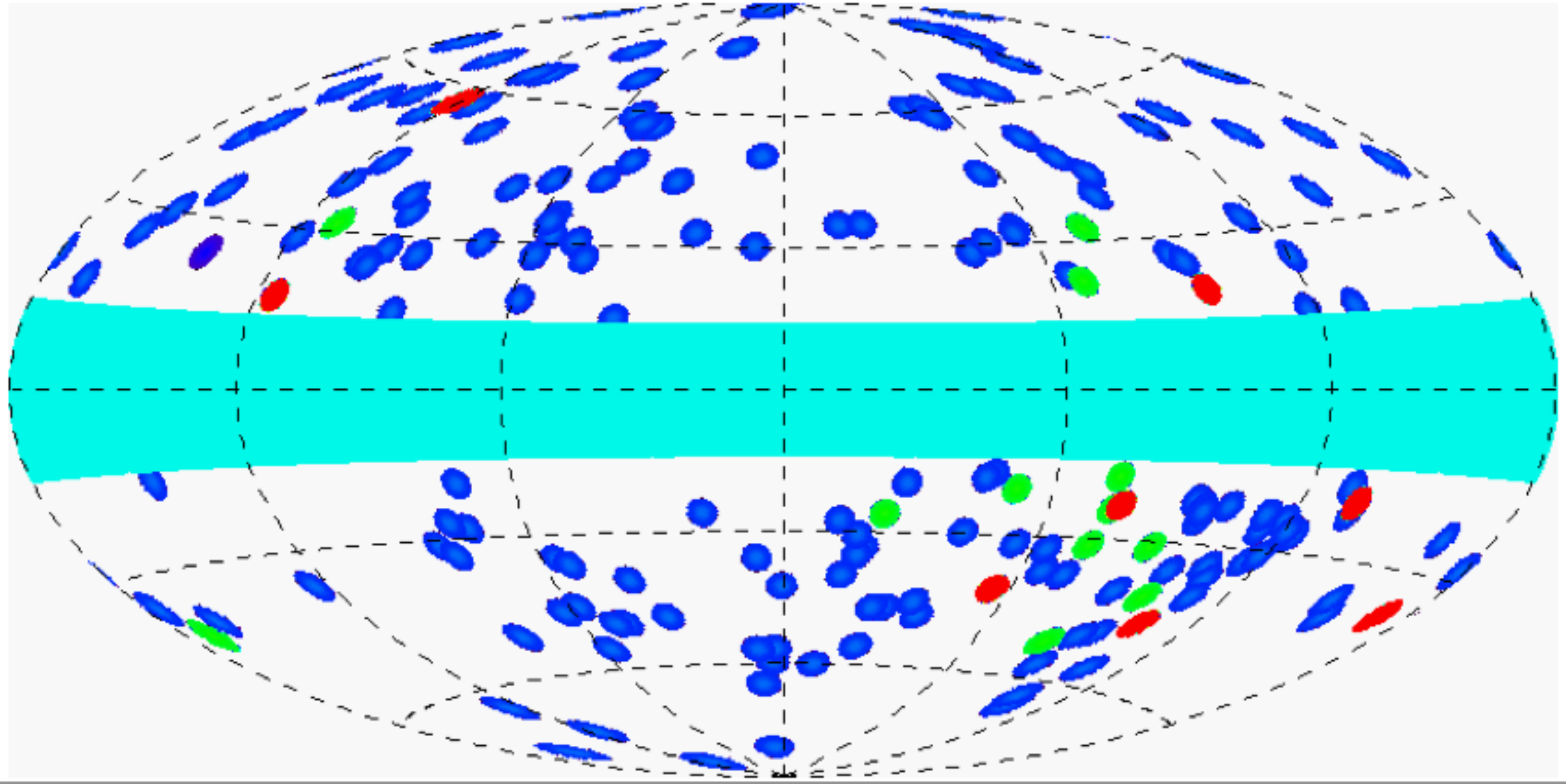}
\end{center}
\caption{Fig. 3 from \cite{PlanckClusterCatalog_Paper8}. Distribution of {\it Planck} clusters that were already known
(blue) and the new cluster candidates (green and red). Credit: {\it Planck} Collaboration, A\&A, Vol. 536, A8,
2011, reproduced with permission by ESO.} \label{F2}
\end{figure}

Among the first results published by the {\it Planck} collaboration on the SZ effect, the Coma cluster (see Fig.
\ref{F1}) constitutes one of the most spectacular ones. Coma is a nearby massive cluster that is well resolved by
{\it Planck}. Fig. \ref{F1} shows the power of {\it Planck} to study the SZ effect with unprecedented quality. In
the near future, studies based on {\it Planck} data alone or combined with X-ray data will reveal new details
about the internal structure of this and other clusters.

\begin{figure}[htb]
\begin{center}
\includegraphics[width=.45\textwidth]{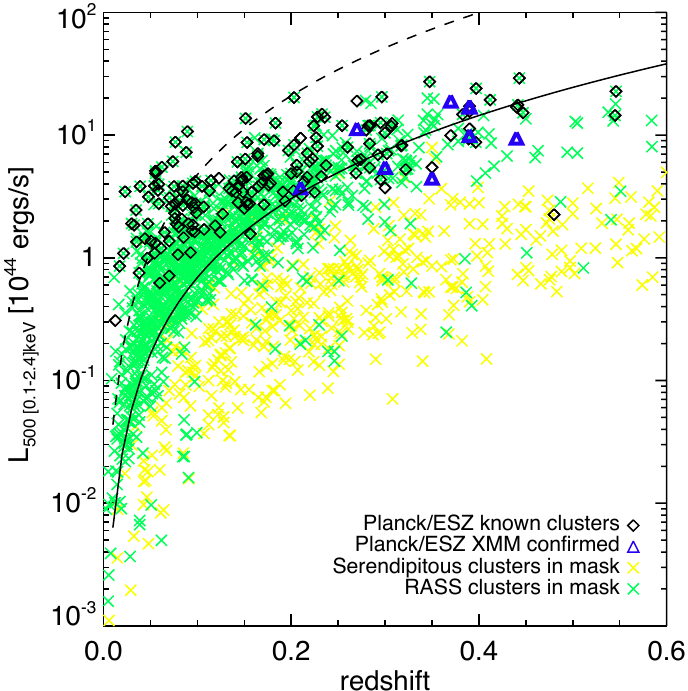}
\end{center}
\caption{Fig. 21 from \cite{PlanckClusterCatalog_Paper8}. Redshift distribution as a function of luminosity for
the 158 clusters from the {\it Planck} Early SZ sample (diamonds and triangles) identified with known X-ray clusters,
compared with serendipitous and RASS clusters (crosses). See \cite{PlanckClusterCatalog_Paper8} for more details.
Credit: {\it Planck} Collaboration, A\&A, Vol. 536, A8, 2011, reproduced with permission by ESO.}
\label{F3}
\end{figure}

{\it Planck}'s earlier results include the detection of almost 200 clusters through their SZ signature
(\cite{PlanckClusterCatalog_Paper8}). {\it Planck} is particularly sensitive to phenomena that increase the
pressure, like mergers or superclusters. One such supercluster was detected by {\it Planck}
\citep{PlanckXMM_Paper9}. Most of the clusters seen by {\it Planck} in the early analysis were known nearby
objects but some of them were new clusters, that have been later confirmed by X-ray and/or optical follows up.
Fig. \ref{F2} shows the distribution in the sky of clusters of galaxies as seen by {\it Planck}. This includes the
most massive clusters in the nearby and intermediate distance Universe. The redshift independence of the SZ effect
can be appreciated in Fig. \ref{F3} (next page), which shows the relative flatness of the selection function of
{\it Planck} as compared to cluster selections based on X-ray luminosity. New analysis based on better data will
improve the selection function by reducing the limiting mass as a function of redshift. It is expected
\cite{BlueBook2005} that {\it Planck} will increase the number of known clusters in a significant way and, in
particular, it will explore the high--redshift regime, detecting the most massive clusters at these high
redshifts.

\begin{figure}
\begin{center}
\includegraphics[width=.45\textwidth]{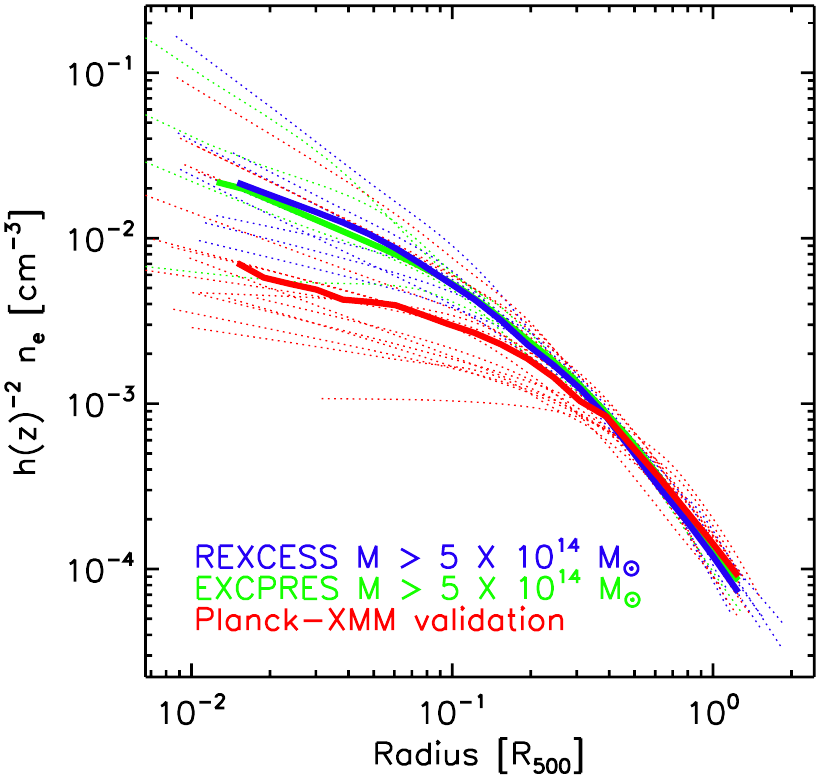}
\end{center}
\caption{Fig. 10 from \cite{PlanckXMM_Paper9}. Scaled density profiles, derived from X-ray data, of the new {\it
Planck} SZ clusters compared to those of similar mass systems from representative X--ray samples. Thick lines show
the mean profile of each sample. Credit: {\it Planck} Collaboration, A\&A, Vol. 536, A9, 2011, reproduced with
permission by ESO.} \label{F4}
\end{figure}

One of the most interesting conclusions derived by these earlier results comes from the combination of X-ray and
SZ data. Earlier studies based on X-ray data were able to conclude that there exists a universal profile that
accurately describes the gas pressure in galaxy clusters \citep{Arnaud_UP}. The newly discovered (by {\it Planck})
SZ clusters seem to follow well this profile but small deviations were observed when comparing the mean SZ profile
with the average profile derived from X-ray observations. Fig. \ref{F4} summarizes one of the main results of the
paper \cite{PlanckXMM_Paper9} where it can be appreciated how the average profile of the SZ observations (red
thick line) flattens towards the cluster center ($R_{500}\ll 1$) when compared to the average of a sample of cool,
core relaxed X-ray clusters (thick blue line). This fact suggests that the new clusters detected by {\it Planck}
tend to be non--cool core, morphologically disturbed clusters. This would explain why these clusters where not
found by previous X-ray surveys but {\it Planck}, that is sensitive to the total pressure rather than to the
distribution of the gas, has no problem in finding them.

\begin{figure}[htb]
\begin{center}
\includegraphics[width=.9\textwidth]{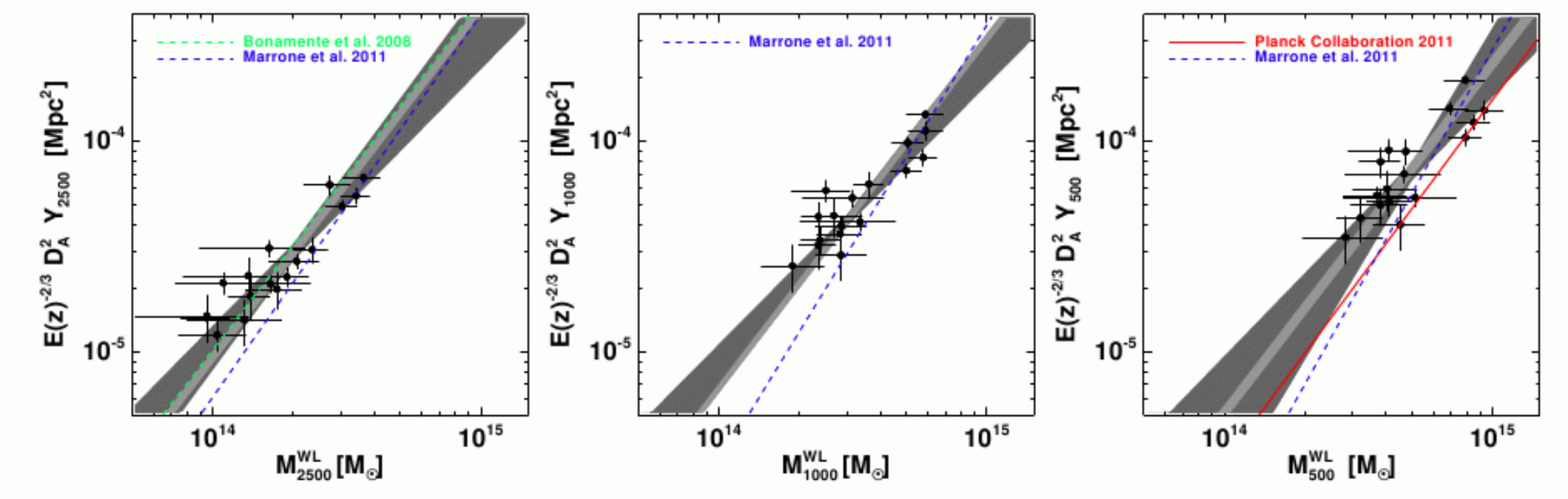}
\end{center}
\caption{Fig. 1 from \cite{Planck_paperWL}. Correlation between the Compton parameter and the weak lensing mass.
Credit: {\it Planck} Collaboration, A\&A, submitted (ms AA/2012/19398), 2012, reproduced with permission
by ESO.} \label{F5}
\end{figure}

Many other relevant results can be found in the first series of papers from the {\it Planck} ERCSC, including
studies of scaling relations between SZ quantities and optical or X-ray ones. More recently, a new analysis based
on 2.5 full sky surveys has studied the relationship between the Compton parameter and weak lensing mass estimates
\cite{Planck_paperWL}. As shown in Fig. \ref{F5}, this work is very promising and could allow -- in the near
future -- the use of the Compton parameter as a mass proxy in cosmological studies, where a good mass estimator is
crucial to derive accurate cosmological parameters from the analysis of a cluster sample.

After {\it Planck}'s data release, science based on the SZ effect will change dramatically. {\it Planck} is
expected to release more than 2 full sky surveys of data early in 2013, thus opening the door for multiple studies
to be carried out by the scientific community. Cluster science based on the SZ effect will motivate many of these
studies. Of particular interest will be those works that combine SZ effect and X-ray data. The different
dependence of the SZ effect and X-ray emission with electron density and temperature allows for deprojection
techniques to reduce the uncertainties of the models. Also, the combination of X-ray and SZ data can be
particularly powerful to study the clumpiness of the gas and deviations from spherical symmetry. An area where
future {\it Planck} data will be used extensively will be the detection of new cluster candidates. The legacy {\it
Planck} cluster catalogue will contain the most significant cluster signals. Hundreds of weaker SZ (and unknown)
clusters will still be present in the public data but not in the legacy catalogue. Many groups will dig into the
{\it Planck} data searching for these weaker signals. Among them there
will be the most distant clusters at $z>1$ that will be crucial for future cosmological studies. \\

Another area where {\it Planck} data might contribute significantly is in the study of energetic phenomena in
galaxy clusters. The SZ effect is sensitive to the temperature of the Plasma (or more generally, to the speed
distribution of the electrons). Hot clusters have an SZ spectrum that deviates from the standard shape. The shift
(or relativistic correction) is more dramatic at higher frequencies ($\nu > 100$ GHz). Current X-ray missions like
{\it Chandra} have some trouble determining the temperature of the plasma for clusters with high temperatures. On
the contrary, the relativistic corrections to the SZ effect can dramatically boost the SZ signal in {\it Planck}
at frequencies $\nu > 500$ GHz making, it easier to detect these clusters at these frequencies and also to  derive
constraints on the physical state of the plasma. A strong deviation in the spectrum could be an indication that
very energetic phenomena (very high temperatures, shock waves, etc.) are operating on the cluster at large scales.

\section{Extragalactic radio and far--IR sources}

The {\it Planck} ERCSC \cite{Planck_Paper7} provides positions and flux densities of compact sources found in each
of the nine {\it Planck} frequency maps. The flux densities are calculated using aperture photometry, with careful
modeling of {\it Planck'\rm{s}} elliptical beams\footnote{Flux densities taken from the ERCSC should be divided by
the appropriate colour correction to give the correct flux values for an assumed narrow band measurement at the
central frequency.}. These data on sources detected during the first 1.6 full-sky surveys offers, among other
things, the opportunity of studying the statistical and emission properties of extragalactic radio and far--IR
sources over a broad frequency range, never before fully explored by blind surveys.

As shown  by \cite{Planck_Paper7}, their Table 1, the full-sky surveys of the {\it Planck} satellite are -- and
will be, for years to come -- unique in the millimetre, at $\lambda \leq 3$ mm, and submillimetre domains. The
lack of data in this frequency range represented the largest remaining gap in our knowledge of bright
extragalactic sources (i.e., normal and star--forming galaxies and Active Galactic Nuclei, AGNs) across the
electromagnetic spectrum.
In the course of its planned surveys, {\it Planck} has been able to measure the integrated flux of many hundreds
of ``radio'' sources -- i.e., sources at intermediate to high--redshift, dominated by synchrotron emission due to
hot electrons in the inner jets of the Active Galactic Nucleus (AGN) of the source -- and of many thousands
```far--IR'' sources -- i.e., low--redshift dusty galaxies or sources with emission dominated by interstellar dust
in thermal equilibrium with the radiation field -- thus providing the {\it first complete full-sky catalogue}
(ERCSC) of bright submillimetre sources. Thanks to this huge amount of new data it is thus possible to investigate
the spectral energy distributions (SEDs) of EPS in a spectral domain very poorly explored before and, at the same
time, their cosmological evolution, at least for some relevant source populations.

\subsection{Synchrotron sources: ``blazars''}

The most recent estimates on source number counts of radio (synchrotron) sources up to $\sim50-70$ \,GHz, and the
optical identifications of the corresponding point sources (see, e.g., \cite{Massardi08}), show that these counts
are dominated by radio sources whose average spectral index is ``flat'', i.e., $\alpha\simeq 0.0$ (with the usual
convention $S_\nu\propto\nu^\alpha$). This result confirms that the underlying source population is essentially
made of Flat Spectrum Radio Quasars (FSRQ) and BL Lac objects, collectively called ``blazars''\footnote{Blazars
are jet-dominated extragalactic objects characterized by a strongly variable and polarized emission of the
non-thermal radiation, from low radio energies up to high energy gamma rays \cite{UrryPadovani95}.}, with minor
contributions coming from other source populations \citep{Toffolatti98,deZotti05}. At frequencies $> 100$\,GHz,
however, there is now new information for sources with flux densities below about $1\,$Jy, coming from the South
Pole Telescope (SPT) collaboration \citep{Vieira10}, with surveys over 87 deg$^2$ at 150 and 220\,GHz, and from
the Atacama Cosmology Telescope (ACT) survey over 455 deg$^2$ at 148\,GHz \citep{Marriage11}.

The ``flat'' spectra of blazars are generally believed to result from the superposition of different components in
the inner part of AGN relativistic jets, each with a different synchrotron self-absorption frequency
\cite{KellermannPauliny-Toth69}. At a given frequency, the observed flux density is thus dominated by the
synchrotron-emitting component which becomes self-absorbed, and, in the equipartition regime, the resulting
spectrum is approximately flat.
Given their sensitivity and full sky coverage, {\it Planck} surveys
are uniquely able to shed light on this transition from an almost
``flat'' to a ``steep'' regime in the spectra of blazar sources,
which can be very informative on the ages of sources and on the
inner jet processes which determine the acceleration of the
electrons \cite{Marscher80}.

To study the spectral properties of the extragalactic radio sources in the {\it Planck} ERCSC
\cite{Planck_Paper13} used a reference 30\,GHz sample above an estimated completeness limit $S_{lim}\simeq
1.0\,$Jy. Not all of these sources were detected at the $\ge 5 \sigma $ level in each of the {\it Planck}
frequency channels considered. Whenever a source was not detected in a given channel they replaced its (unknown)
flux density by a $5\sigma$ upper limit, where for $\sigma$ they used the average r.m.s. error estimated at each
{\it Planck} frequency. Finally, for estimating spectral index distributions, these upper limits have been
redistributed among the flux density bins by using a Survival Analysis technique and, more specifically, by
adopting the Kaplan-Meyer estimator\footnote{Since the fraction of upper limits is found to be always small (it
reaches approximately $30\%$ only in the less sensitive channel at 44GHz), the spectral index distributions are
reliably reconstructed at each frequency.}\cite{Planck_Paper13}.

\begin{figure}[htb]
\begin{center}
\includegraphics[width=.45\textwidth]{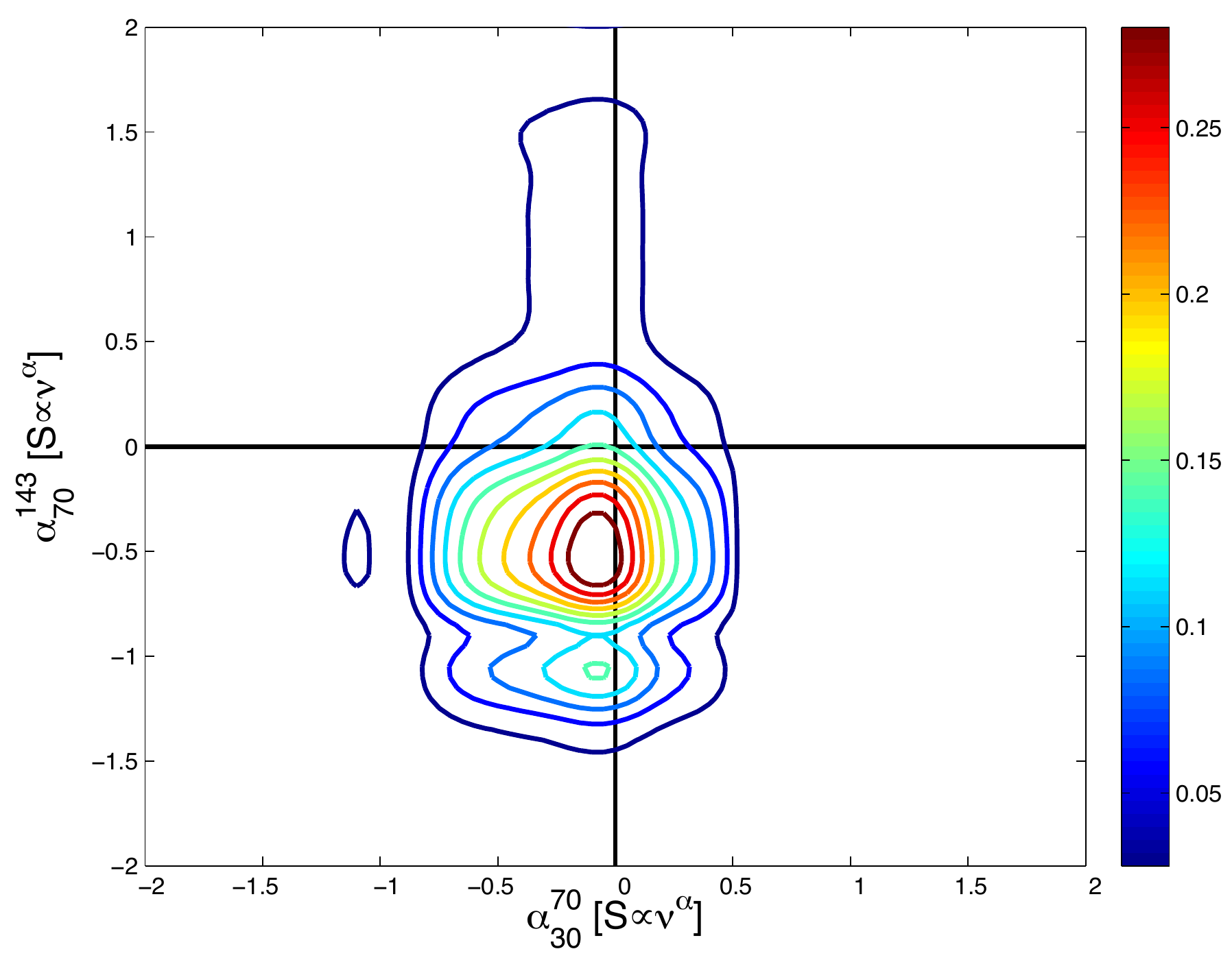}
\end{center}
\caption{Fig. 7 from \cite{Planck_Paper13}. Contour levels of the distribution of ${\alpha}_{70}^{143}$ vs.
$\alpha_{30}^{70}$ obtained by Survival Analysis, i.e., taking into account the upper limits to flux densities at
each frequency. The colour scale can be interpreted as the probability of having any particular pair of values of
the two spectral indices. The maximum probability corresponds to $\alpha_{30}^{70}\simeq -0.18$ and
$\alpha_{70}^{143}\simeq -0.5$. Credit: {\it Planck} Collaboration, A\&A, Vol. 536, A13, 2011, reproduced with
permission by ESO.} \label{Two_colours}
\end{figure}

In the sample analyzed by \cite{Planck_Paper13}, the 30--100\,GHz median spectral index is very close to the
$\alpha\simeq -0.39$ found by \cite{Sadler08} between 20 and 95\,GHz, for a sample with 20\,GHz flux density
$S>150\,$mJy. Moreover, the 30--143\,GHz median spectral index is in very good agreement with the one found by
\cite{Marriage11} for their bright ($S_\nu>50$ mJy) 148\,GHz-selected sample with complete cross-identifications
from the Australia Telescope 20\,GHz survey, i.e $\alpha_{20}^{148}=-0.39\pm 0.04$. Fig.\ref{Two_colours} presents
the contour levels of the distribution of $\alpha_{70}^{143}$ vs. $\alpha_{30}^{70}$ (obtained using Survival
Analysis) in the form of a 2D probability field: the colour scale can be interpreted as the probability of a given
pair of spectral indices and a bending down, i.e. $\alpha< -0.5$, at high frequencies is displayed. In the whole,
the results of \cite{Planck_Paper13} show that in their sample selected at 30 GHz a moderate steepening of
spectral indices of EPS at high radio frequencies, i.e. $\gsim 70-100$ GHz, is clearly apparent\footnote{Some
hints in this direction were previously found by other works on the subject. Additional evidence of spectral
steepening is also presented in \cite{Planck_Paper15} by the analysis of a complete sample of blazars selected at
37\,GHz.}.

As already noted, at high radio frequencies ($\nu > 30$ GHz) most of the bright extragalactic radio-sources are
blazars. From the contour plot of Fig.~\ref{Two_colours} it is possible to see that the maximum probability of the
spectral indices of blazars corresponds to $\alpha^{70}_{30}\simeq -0.18$ and $\alpha^{143}_{70} \simeq -0.5$. A
secondary maximum can also be seen at $\alpha^{143}_{70} \simeq -1.2$. In a companion paper, i.e.
\cite{Planck_Paper15}, a detailed discussion on the modelling of the spectra of this source class is also
presented. In this paper, spectral energy distributions (SEDs) and radio continuum spectra are presented for a
northern sample of 104 extragalactic radio sources, based on the {\it Planck} ERCSC and simultaneous
multifrequency data\footnote{The great amount of data present in the {\it Planck} ERCSC complemented with
quasi-simultaneous ground--based observations at mm wavelengths have also enabled the study of the very
interesting spectral properties of the rare peculiar and/or extreme radio sources detected by the {\it Planck}
surveys \cite{Planck_Paper14}.}. The nine {\it Planck} frequencies, from 30 to 857 GHz, are complemented by a set
of quasi--simultaneous observations ranging from radio to gamma-rays. SED modelling methods are discussed, with an
emphasis on proper, physical modelling of the synchrotron bump using multiple components, and a thorough
discussion on the original accelerated electron energy spectrum in blazar jets is presented. The main conclusion
is that, al least for a fraction of the observed mm/sub-mm blazar spectra, the energy spectrum could be much
harder than commonly thought, with a power-law index $\sim 1.5$ and the implications of this hard value are
discussed for the acceleration mechanisms effective in blazar shocks.

It has also been shown by \cite{Planck_Paper13} that differential number counts at 30, 44, and 70\,GHz are in good
agreement with those derived from \textit{WMAP} \cite{wright09} data at nearby frequencies. 
The model proposed by \cite{deZotti05} is consistent with the present counts at frequencies up to 70\,GHz, but
over-predicts the counts at higher frequencies by a factor of about 2.0 at 143\,GHz and about 2.6 at
217\,GHz\footnote{This implies that the contamination of the CMB APS by radio sources below the 1\,Jy detection
limit is lower than previously estimated. No significant changes are found, however, if we consider fainter source
detection limits, i.e., 100\,mJy, given the convergence between predicted and observed number counts.}. As shown
above, the analysis of the spectral index distribution over different frequency intervals, within the uniquely
broad range covered by {\it Planck} in the mm and sub-mm domain, has highlighted an average {\it
steepening} of source spectra above about 70\,GHz. 
This steepening accounts for the discrepancy between the model predictions of \cite{deZotti05} and the observed
differential number counts at HFI frequencies.

In the fall of 2011, a successful explanation of the change detected in the spectral behavior of extragalactic
radio sources (ERS) at frequencies above 70-80\,GHz has been proposed by \cite{Tucci11}. This paper makes a first
attempt at constraining the most relevant physical parameters that characterize the emission of blazar sources by
using the number counts and the spectral properties of extragalactic radio sources estimated from high--frequency
radio surveys\footnote{The main goal of \cite{Tucci11} was to present physically grounded models to extrapolate
the number counts of ERS, observationally determined over very large flux density intervals at cm wavelengths down
to mm wavelengths, where experiments aimed at accurately measuring CMB anisotropies are carried out.}. As noted
before, a relevant steepening in blazar spectra with emerging spectral indices in the interval between $-0.5$ and
$-1.2$, is commonly observed at mm/sub-mm wavelengths. \cite{Tucci11} interpreted this spectral behavior as
caused, at least partially, by the transition from the optically--thick to the optically--thin regime in the
observed synchrotron emission of AGN jets \cite{Marscher96}. Indeed, a ``break'' in the synchrotron spectrum of
blazars, above which the spectrum bends down, thus becoming ``steep'', is predicted by models of synchrotron
emission from inhomogeneous unresolved relativistic jets \cite{Blandford79,Konigl81}. Based on these models,
\cite{Tucci11} estimated the value of the frequency $\nu_M$ (and of the corresponding radius $r_M$) at which the
break occurs on the basis of the flux densities of ERS measured at 5\,GHz and of the most typical values for the
relevant physical parameters of AGNs.

\begin{figure}[htb]
\begin{minipage}[b]{0.48\linewidth}
\centering
\includegraphics[width=1.02\textwidth]{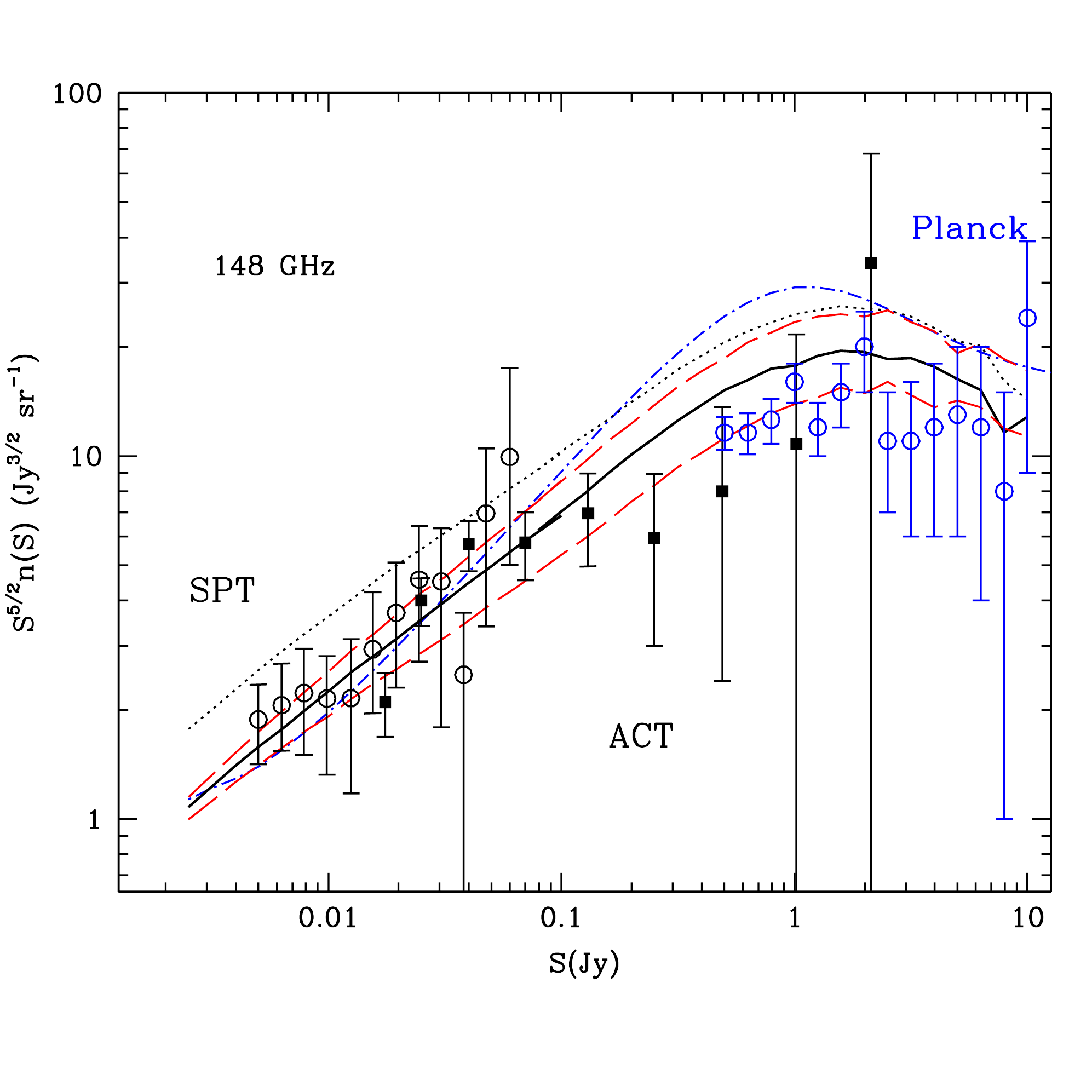}
\end{minipage}
\begin{minipage}[b]{0.48\linewidth}
\centering
\includegraphics[width=1.02\textwidth]{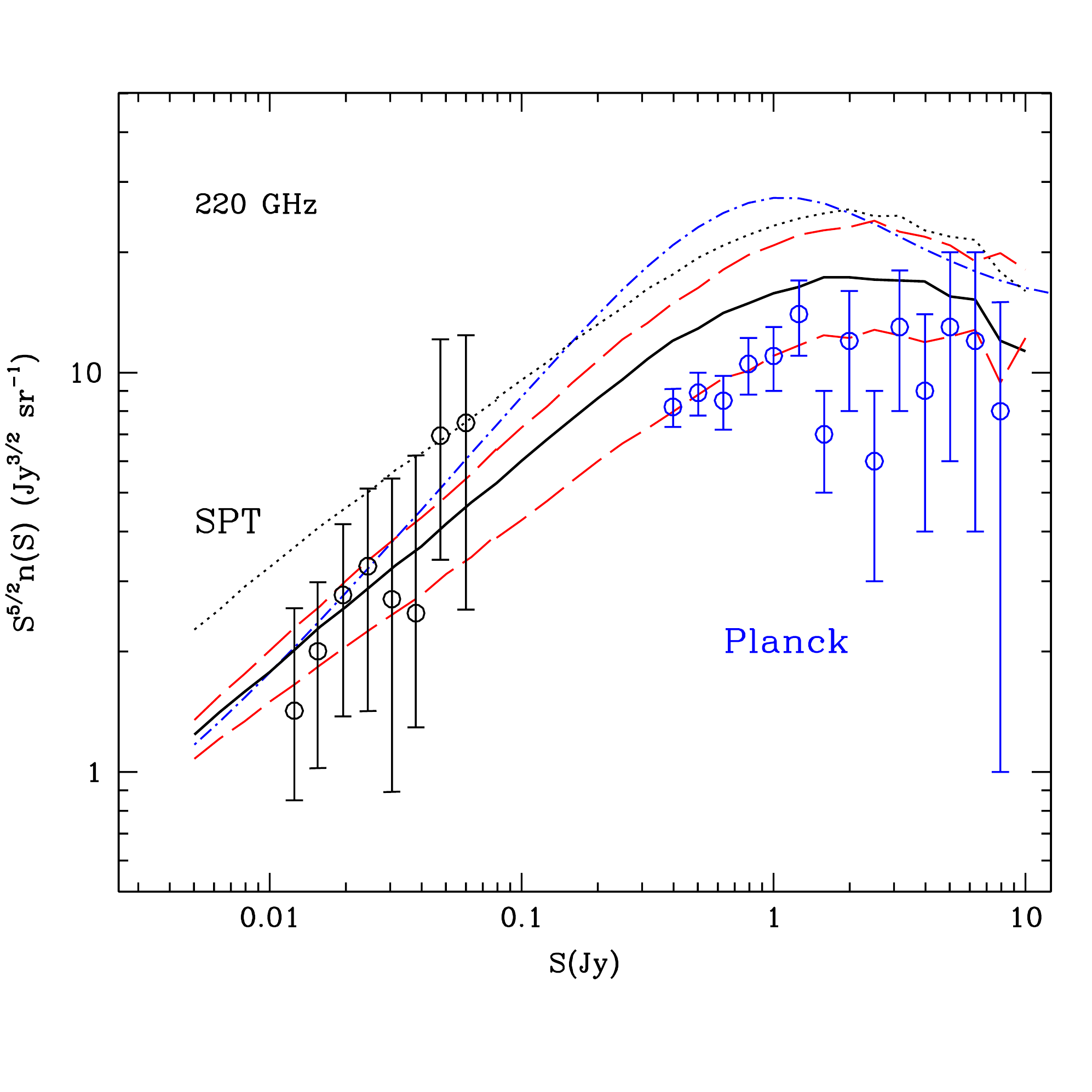}
\end{minipage}
\vskip -0.6cm \caption{Comparison between predicted and observed
differential number
  counts at 148\,GHz ({\it left panel}) and at 220\,GHz ({\it right
    panel}). Filled circles: ACT data; open black circles: SPT data;
  open blue circles: {\it Planck} ERCSC counts \cite{Planck_Paper13} at
  143\,GHz (left panel) and 217\,GHz (right panel). The plotted
  lines indicate predictions of different models, as
  follows: {\bf C0}(dotted lines), {\bf C1} (thick continuous lines), {\bf
    C2Ex} (lower red long--dashed lines) and {\bf C2Co} (upper red
  long--dashed lines) and the \cite{deZotti05} model (blue dash--dotted
  line). Credit: Tucci M., et al., A\&A, Vol. 533, A57, 2011,
  reproduced with permission by ESO.}
\label{F10}
\end{figure}

As displayed in Fig. \ref{F10}, high frequency ($\nu \ge 100\,$GHz) data on source number counts are the most
powerful for distinguishing among different cosmological evolution models (see \cite{Tucci11} for more details on
the models plotted in Fig. \ref{F10})\footnote{The two most relevant models of \cite{Tucci11}, i.e. {\bf C2Co} and
{\bf C2Ex}, assume different distributions of $r_M$ -- i.e., the smallest radius in the AGN jet from which
optically-thin synchrotron emission can be observed -- for BL\,Lacs and FSRQs, with the former objects that
generate, in general, the synchrotron emission from more compact regions, implying higher values of $\nu_M$ (above
100\,GHz for bright objects). These two models differ only in the $r_M$ distributions for FSRQs: in the {\bf C2Co}
model the emitting regions are more compact, implying values of $\nu_M$ partially overlapping with those for
BL\,Lacs, whereas in the {\bf C2Ex} model they are more extended, thus predicting very different values of $\nu_M$
for FSRQs and BL\,Lacs.}. As clearly shown, these most recent data on number counts require spectral ``breaks'' in
blazars' spectra and clearly favor the model {\bf C2Ex}. According to this, most of the FSRQs (which are the
dominant population at low frequencies and at Jy flux densities), differently from BL\,Lacs, should bend their
flat spectrum before or around 100\,GHz. The {\bf C2Ex} model also predicts a substantial increase of the BL\,Lac
fraction at high frequencies and bright flux densities\footnote{This is indeed observed: a clear dichotomy between
FSRQs and BL\,Lac objects has been found in the {\it Planck} ERCSC. Almost all radio sources show very flat
spectral indices at LFI frequencies, i.e. $\alpha_{LFI}\geq -0.2$, whereas at HFI frequencies, BL\,Lacs keep flat
spectra, i.e. $\alpha_{HFI}\geq -0.5$, with a high fraction of FSRQs showing steeper spectra, i.e.
$\alpha_{HFI}\leq -0.5$.}. On the whole, the results of \cite{Tucci11} imply that the parameter $r_M$ should be of
parsec--scales, at least for FSRQs, in agreement with the theoretical predictions of \cite{Marscher85}, whereas
values of $r_M\ll 1\,$pc should be only typical of BL\,Lac objects or of rare, and compact, quasar sources.

\subsection{Far--IR sources: local dusty galaxies}

The full-sky coverage of the {\it Planck} ERCSC provides an unsurpassed survey of galaxies at submillimetre
(submm) wavelengths, representing a major improvement in the numbers of galaxies detected, as well as the range of
far-IR/submm wavelengths over which they have been observed. The analysis done by \cite{Planck_Paper16} presented
the first results on the properties of nearby galaxies using these data. They matched the ERCSC catalogue to
IRAS-detected galaxies in the Imperial IRAS Faint Source Redshift Catalogue (IIFSCz) \cite{Wang09}, so that they
could measure the SEDs of these objects from 60 to 850 $\mu$m. This produced a list of 1717 galaxies with reliable
associations between {\it Planck} and IRAS, from which they selected a subset of 468 for SED studies, namely those
with strong detections in the three highest frequency {\it Planck} bands and no evidence of cirrus contamination. This
selection has thus provided a first {\it Planck} sample of local, i.e. at redshift $< 0.1$, dusty galaxies, very
important for determining their emission properties and, in particular, the presence of different dust components
contributing to their submm SEDs. Moreover, the richness of data on extragalactic point sources gathered by {\it
Planck} has allowed the measurement of the submm number density of bright ($S > 0.5-2$ Jy) dusty galaxies (and of
synchrotron-dominated sources) for the first time.

\begin{figure}[htb]
\begin{center}
\includegraphics[width=.8\textwidth]{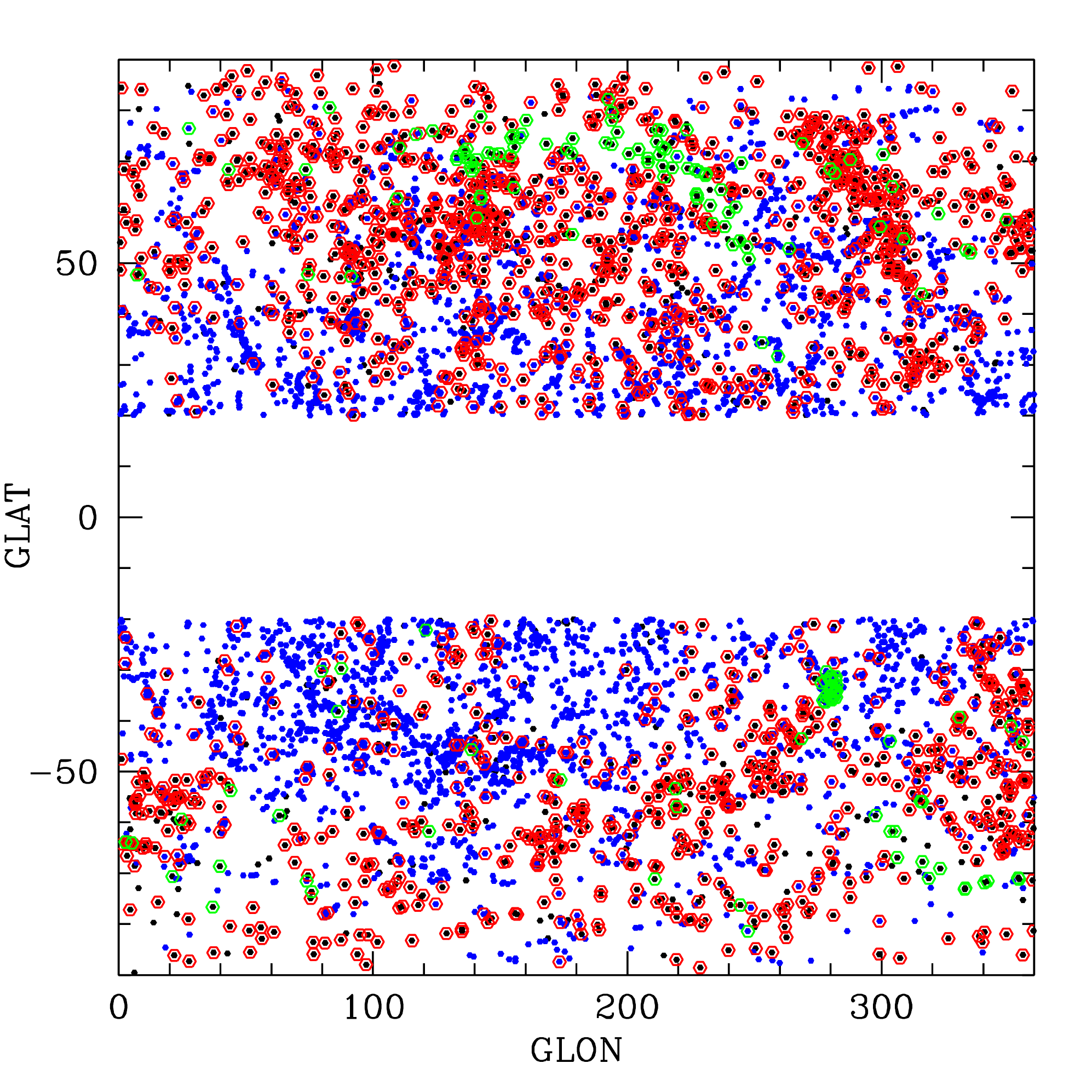}
\end{center}
\vskip -0.4cm \caption{Fig. 2 from \cite{Planck_Paper16}. Sky plot of ERCSC sources in Galactic coordinates. ERCSC
point-sources (black filled hexagons) and ERCSC sources flagged as extended (blue filled hexagons) are shown. Red
hexagons are sources associated with IIFSCz IRAS FSC galaxies. Green hexagons are ERCSC sources not associated
with IIFSCz, but associated with bright galaxies in NED (only for $\vert b\vert > 60^\circ$ for extended sources).
Credit: {\it Planck} Collaboration, A\&A, Vol. 536, A16, 2011, reproduced with permission by ESO.} \label{F11}
\end{figure}

Fig. \ref{F11} shows the sky distribution of ERCSC sources at $\vert b\vert > 20^\circ$, with sources flagged as
extended in the ERCSC shown as blue filled hexagons, and point-sources shown in black. Associations with the
IIFSCz are shown as red circles. The extended sources not associated with IIFSCz sources have a strikingly
clustered distribution, which matches the areas of our Galaxy with strong cirrus emission, as evidenced by IRAS
100 $\mu$m maps. Therefore, the majority of these are cirrus sources and not extragalactic (see
\cite{Planck_Paper16} for more details).

The studies of nearby galaxies detected by {\it Planck} \cite{Planck_Paper16} confirm the presence of cold dust in
local galaxies and also largely in dwarf galaxies. The SEDs are fitted using parametric dust models to determine
the range of dust temperatures and emissivities. They found evidence for colder dust than has previously been
found in external galaxies, with temperatures $T < 20$ K. Such cold temperatures are found by using both the
standard single temperature dust model with variable emissivity $\beta$, or a two dust temperature model with
$\beta$ fixed at 2. In \cite{Planck_Paper16} it is also found that some local galaxies are both luminous and cool,
with properties similar to those of the distant submm galaxies uncovered in deep surveys. This suggests that
previous studies of dust in local galaxies have been biased away from such luminous cool objects. In most galaxies
the dust SEDs are found to be better described by parametric models containing two dust components, one warm and
one cold, with the cold component reaching temperatures as low as $10$ K\footnote{Fits to SEDs of selected objects
using more sophisticated templates derived from radiative transfer models confirm the presence of the colder dust
found through parametric fitting.}. The main conclusion of \cite{Planck_Paper16} is that cold ($T < 20$ K) dust is
thus a significant and largely unexplored component of many nearby galaxies. Furthermore, a new population of cool
submm galaxies is detected, with presence of very cold dust ($T= 10-13$ K) showing a more extended spatial
distribution than generally assumed for the gas and dust in galaxies.

\begin{figure}[htb]
\begin{center}
\includegraphics[width=.9\textwidth]{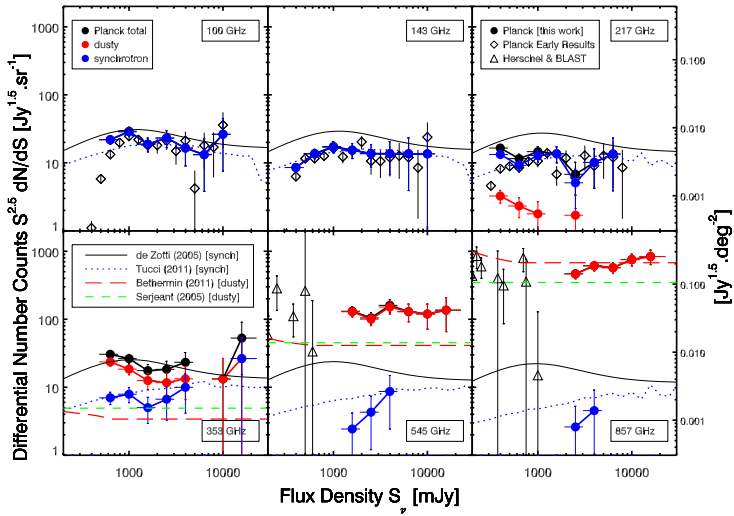}
\end{center}
\vskip -6.5cm \caption{Fig. 9 from \cite{Planck_IntPap7}. {\it Planck}
   differential number counts, normalized to the
    Euclidean value (i.e.\ $S^{2.5} dN/dS$), compared with models and
    other data sets. {\it Planck} counts: total (black filled circles);
    dusty (red circles); synchrotron (blue circles). Four models are
    also plotted: \cite{deZotti05}, dealing only with synchrotron sources -- solid line;
    \cite{Tucci11} dealing only with synchrotron sources -- dots;
    \cite{Bethermin11} dealing only with dusty sources -- long dashes;
    \cite{Serjeant05} dealing only with local dusty sources -- short dashes.
    Other data sets:
    {\it Planck} early counts for 30\,GHz-selected radio galaxies
    \cite{Planck_Paper13} at 100, 143 and 217\,GHz (open diamonds);
    {\it Herschel} ATLAS and HerMES counts at 350 and 500 micron from
    \cite{Oliver10} and \cite{Clements10}; BLAST at the same two
    wavelengths, from \cite{Bethermin10}, all shown as
    triangles. Left vertical axes are in units of
    Jy$^{1.5}$\,sr$^{-1}$, and the right vertical axis in
    Jy$^{1.5}$.deg$^{-2}$. Credit: {\it Planck} Collaboration, A\&A, submitted (ms AA/2012/20053), 2012, reproduced with permission by ESO.} \label{F12}
\end{figure}

Very recently, using EPS samples selected from the first {\it Planck} 1.6 full-sky surveys, i.e. from the {\it
Planck} ERCSC, \cite{Planck_IntPap7} have derived number counts of extragalactic sources from 100 to 857\,GHz
(3\,mm to 350 $\mu$m). Three zones (deep, medium and shallow) of approximately homogeneous coverage are used to
ensure a clean completeness correction\footnote{The sample, prior to the 80\,\% completeness cut, contains between
217 sources at 100\,GHz and 1058 sources at 857\,GHz over about 12,800 to 16,550\,deg$^2$ (31 to 40\,\% of the
sky). After the 80\,\% completeness cut, between 122 and 452 and sources remain, with flux densities above 0.3 and
1.9\,Jy, at 100 and 857 GHz, respectively.}. For the first time, bright number counts of EPS at 353, 545 and
857\,GHz (i.e., 850, 550 and 350 $\mu$m) have been calculated\footnote{More specifically, number counts have been
provided of synchrotron-dominated sources at high frequency (353 to 857\,GHz) and of dusty-dominated galaxies at
lower frequencies (217 and 353\,GHz).}. {\it Planck} number counts are found to be in the Euclidean regime in this
frequency range, since the ERCSC comprises only bright sources ($S> 0.3$ Jy). The estimated number counts appear
generally in agreement with other data sets, when available (see \cite{Planck_IntPap7} for more details).

Using multi-frequency information to classify the sources as dusty- or synchrotron-dominated (and measure their
spectral indices), the most striking result of \cite{Planck_IntPap7} is the estimated contribution to the number
counts by each population. 
These new estimates of number counts of synchrotron and of dust--dominated EPS (displayed in Fig.~\ref{F12}) have
allowed new constraints to be placed on models which extend their predictions to bright flux densities, i.e. $S>
1$ Jy. A very relevant result is that the model {\bf C2Ex} of \cite{Tucci11} (see Section 5.1) is performing
particularly well at reproducing the number counts of synchrotron-dominated sources up to 545\,GHz. On the
contrary, \cite{Planck_IntPap7} highlights the failure of many models for number count predictions of dusty
sources to reproduce all the high-frequency counts. The model of \cite{Bethermin11} agrees marginally at 857\,GHz
but is too low at 545\,GHz and also at lower frequencies, whereas the model of \cite{Serjeant05} is marginally
lower at 857\,GHz, fits the data well at 545\,GHz, but is too low at 353\,GHz. The likely origin of the
discrepancies is an inaccurate description of the galaxy SEDs used at low redshift in these models. Indeed a cold
dust component, detected by \cite{Planck_Paper16}, is rarely included in the models of galaxy SEDs at low
redshift. On the whole, these results already obtained by the exploitation of the {\it Planck} ERCSC data are
providing valuable information about the ubiquity of cold dust in the local Universe, at least in statistical
terms, and are guiding to a better understanding of the cosmological evolution of EPS at mm/sub-mm wavelengths.

\section{Cosmic Infrared Background anisotropies}

The Cosmic Infrared Background (CIB) is the relic emission, at wavelengths larger than a few microns, of the
formation and evolution of the galaxies of all types, including AGNs and star-forming systems
\cite{Puget96,Hauser01,Dole06}\footnote{An important goal of studies of galaxy formation has thus been the
characterization of the statistical behavior of galaxies responsible for the CIB - such as the number counts,
redshift distribution, mean SED, luminosity function, clustering -- and their physical properties, such as the
roles of star-forming vs. accreting systems, the density of star formation, and the number density of very hot
stars}. The CIB accounts for roughly half of the total energy in the optical/infrared Extragalactic Background
Light (EBL) \cite{Hauser01}, although with some uncertainty, and its SED peaks near 150 $\mu$m. Since local
galaxies give rise to an integrated infrared output that amounts to only about a third of the optical one
\cite{Soifer91}, there must have been a strong evolution of galaxy properties towards enhanced far--IR output in
the past. Therefore, the CIB, made up by high density, faint and distant galaxies\footnote{The CIB records much of
the radiant energy released by processes of structure formation occurred since the last scattering epoch, four
hundred thousand years after the Big Bang, when the CMB was produced.} is barely resolved into its constituents.
Indeed, less than 10\% of the CIB is resolved by the {\it Spitzer} satellite at 160 $\mu$m \cite{Bethermin10},
about 10\% by {\sl Herschel} at 350 $\mu$m \cite{Oliver10}. Thus, in the absence of foreground (Galactic dust) and
CMB emissions, and when the instrument noise is subdominant, maps of the diffuse emission at the angular
resolution probed by the current surveys reveal a web of structures, characteristic of CIB anisotropies. With the
advent of large area far-IR to millimeter surveys ({\sl Herschel}, {\it Planck}, SPT, and ACT), CIB anisotropies
thus constitute a new tool for structure formation and evolution studies.

CIB anisotropies are expected to trace large-scale structures and probe the clustering properties of galaxies,
which in turn are linked to those of their hosting dark matter halos. Because the clustering of dark matter is
well understood, observations of anisotropies in the CIB constrain the relationship between dusty, star-forming
galaxies at high redshift, i.e. $z\geq 2$, and the underlying dark matter distribution.
The angular power spectrum of CIB anisotropies has two contributions, a white-noise component caused by shot noise
and an additional component caused by spatial correlations between the sources of the CIB. Correlated CIB
anisotropies have already been measured by many space--borne as well as ground--based experiments (see
\cite{Planck_Paper18} for more details). Depending on the frequency, the angular resolution and size of the
survey, these measurements can probe two different clustering regimes. On small angular scales ($\ell\geq  2000$),
they measure the clustering within a single dark matter halo and, accordingly, the physics governing how dusty,
star--forming galaxies form within a halo. On large angular scales, i.e. $200\leq\ell\leq 2000$, CIB anisotropies
measure clustering between galaxies in different dark matter halos. These measurements primarily constrain the
large-scale, linear bias, $b$, of dusty galaxies, which is usually assumed to be scale-independent over the
relevant range.


Thanks to the exceptional quality of the {\it Planck} data, \cite{Planck_Paper18} were able to measure the
clustering of dusty, star-forming galaxies at 217, 353, 545, and 857 GHz with unprecedented precision. The CIB
maps were cleaned using templates: HI for Galactic cirrus; and the {\it Planck} 143 GHz maps for CMB. Having HI
data is necessary to cleanly separate CIB and cirrus fluctuations. After careful cleaning, they obtained CIB
anisotropy maps that reveal structures produced by the cumulative emission of high-redshift, dusty, star--forming
galaxies. The maps are highly correlated at high {\it Planck} frequencies, whereas they decorrelate at lower {\it
Planck} HFI frequencies. \cite{Planck_Paper18} then computed the power spectra of the maps and their associated
errors using a dedicated pipeline and ended up with measurements of the APS of the CIB anisotropy, $C_{\ell}$, at
217, 353, 545, and 857 GHz, with high signal-to-noise ratio over the range $200 < l < 2000$. These measurements
compare very well with previous measurements at higher $\ell$\footnote{The SED of CIB anisotropies is not
different from the CIB mean SED, even at 217 GHz. This is expected from the model of \cite{Bethermin11} and
reflects the fact that the CIB intensity and anisotropies are produced by the same population of sources.}.

Moreover, from {\it Planck} data alone \cite{Planck_Paper18} could exclude a model where galaxies trace the
(linear theory) matter power spectrum with a scale-independent bias: that model requires an {\it unrealistic} high
level of shot noise to match the small-scale power they observed. Consequently, an alternative model that couples
the dusty galaxy, parametric evolution model of \cite{Bethermin11} with a halo model approach has been developed
(see \cite{Planck_Paper18}, again, for more details). Characterized by only two parameters, this model provides an
excellent fit to our measured anisotropy angular power spectrum for each frequency treated independently. In the
near future, modelling and interpretation of the CIB anisotropy will be aided by the use of cross-power spectra
between bands, and by the combination of the {\it Planck} and {\it Herschel} data at 857 and 545/600 GHz and {\it
Planck} and SPT/ACT data at 220 GHz.

\section{Acknowledgements}

LT, FA and JMD acknowledge partial financial support from the Spanish Ministry of Science and Innovation (MICINN)
under project AYA2010--21766--C03-01. CB acknowledges partial financial support by ASI through ASI/INAF Agreement
I/072/09/0 for the {\it Planck} LFI Activity of Phase E2 and by MIUR through PRIN 2009 grant n. 2009XZ54H2. The analysis
of the {\it Planck} full-sky maps has been performed by means of the HEALPix package \cite{gorski05}. We thank the
Editorial Board of Astronomy and Astrophysics (European Southern Observatory; ESO) for having granted us the
permission to reproduce many Figures originally published in (or submitted to) the same Journal. Credits are
indicated in each one of the Figures we used. Many thanks are due to Douglas Scott, for carefully reading the
original manuscript and for his very useful suggestions. We also warmly thank the {\it Planck} Collaboration and,
in particular, all the members of the {\it Planck} Working Groups 2, 5 and 6 and of the LFI Core Team A09, with
whom we shared the analysis and the interpretation of {\it Planck} data as for the subject here discussed, i.e.
''extragalactic compact sources''. Finally, we also thank the members of the {\it Planck} Science Team (ST) and of
the {\it Planck} Editorial Board (EB) for granting us the permission of publishing this Chapter.

\section{Authors' information}

Toffolatti Luigi$^{1,2}$, Burigana Carlo$^{3,4}$ , Arg\"ueso Francisco$^{5,2}$, and Diego Jos\'e M.$^2$

1 Department of Physics, University of Oviedo, c. Calvo Sotelo s/n, 33007 Oviedo, Spain

2 IFCA-CSIC, University of Cantabria, avda. los castros s/n, 39005 Santander, Spain;

3 National Institute of Astrophysics - Institute of Space Astrophysics and Cosmic Physics (INAF-IASF), via Piero
Gobetti 101, 40129 Bologna, Italy;

4 Department of Physics, University of Ferrara, via Giuseppe Saragat 1, 44122 Ferrara.

5 Department of Mathematics, University of Oviedo, c. Calvo Sotelo s/n, 33007 Oviedo, Spain

\end{document}